\documentclass[sigconf, nonacm]{acmart}

\AtBeginDocument{%
  }

\copyrightyear{2025}
\acmYear{2025}
\setcctype{by}
\acmConference[CHI '25]{CHI Conference on Human Factors in Computing Systems}{April 26-May 1, 2025}{Yokohama, Japan}
\acmBooktitle{CHI Conference on Human Factors in Computing Systems (CHI '25), April 26-May 1, 2025, Yokohama, Japan}
\acmDOI{10.1145/3706598.3713353}
\acmISBN{979-8-4007-1394-1/25/04}

\usepackage{booktabs}
\usepackage{graphicx}
\usepackage{listings}
\usepackage{xcolor} 
\usepackage{caption}
\usepackage{makecell}
\usepackage{url}
\usepackage{longtable}
\usepackage{color}
\usepackage{tikz}
\usepackage[T1]{fontenc}
\usepackage{lipsum}
\usepackage{stfloats}
\usepackage{tabu} 
\usepackage{tabularx}
\usepackage{multirow}
\usepackage{booktabs}
\usepackage{graphicx}
\usepackage{wrapfig}
\usepackage{hyperref}   
\usepackage{cleveref}
\usepackage{xcolor}
\usepackage{float}
\usepackage{listings} 

\definecolor{PurpleColor}{RGB}{0,0,0}
\newcommand{\RR}[1]{{\color{PurpleColor}#1}}

\definecolor{PinkColor}{RGB}{0, 0, 0}
\newcommand{\RA}[1]{{\color{PinkColor}#1}}

\definecolor{codegreen}{rgb}{0,0.6,0}
\definecolor{codegray}{rgb}{0.5,0.5,0.5}
\definecolor{codepurple}{rgb}{0.58,0,0.82}
\definecolor{backcolour}{rgb}{0.95,0.95,0.92}

\lstdefinestyle{mystyle}{
  backgroundcolor=\color{backcolour}, commentstyle=\color{codegreen},
  keywordstyle=\color{magenta},
  numberstyle=\tiny\color{codegray},
  stringstyle=\color{codepurple},
  basicstyle=\ttfamily\footnotesize,
  breakatwhitespace=false,         
  breaklines=true,                 
  captionpos=b,                    
  keepspaces=true,                 
  numbers=left,                    
  numbersep=5pt,                  
  showspaces=false,                
  showstringspaces=false,
  showtabs=false,                  
  tabsize=2
}
\begin{document}

\title{CPVis: Evidence-based Multimodal Learning Analytics for Evaluation in Collaborative Programming}

\author{Gefei Zhang}
\affiliation{%
  \institution{Zhejiang University of Technology}
  \city{Hangzhou}
  \state{Zhejiang}
  \country{China}}
\email{gefei@zjut.edu.cn}
\orcid{1234-5678-9012}

\author{Shenming Ji}
\affiliation{%
  \institution{Xi'an Jiaotong-Liverpool University}
  \city{Suzhou}
  \state{Jiangsu}
  \country{China}}
\email{shenming.ji21@student.xjtlu.edu.cn}

\author{Yicao Li}
\affiliation{%
  \institution{Zhejiang University of Technology}
  \city{Hangzhou}
  \state{Zhejiang}
  \country{China}}
\email{yicaoli47@gmail.com}

\author{Jingwei Tang}
\affiliation{%
  \institution{Zhejiang University of Technology}
  \city{Hangzhou}
  \state{Zhejiang}
  \country{China}}
\email{jwtang@zjut.edu.cn}

\author{Jihong Ding}
\affiliation{%
 \institution{Hannan University}
 \city{Haikou}
 \state{Hainan}
 \country{China}}
\email{jhding@hainanu.edu.cn}

\author{Meng Xia}
\affiliation{%
  \institution{Texas A\&M University}
  \city{College Station}
  \state{Texas}
  \country{USA}}
\email{mengxia@tamu.edu}

\author{Guodao Sun}
\affiliation{%
  \institution{Zhejiang University of Technology}
  \city{Hangzhou}
  \state{Zhejiang}
  \country{China}}
\email{guodao@zjut.edu.cn}

\author{Ronghua Liang}
\affiliation{%
  \institution{Zhejiang University of Technology}
  \city{Hangzhou}
  \state{Zhejiang}
  \country{China}}
\email{rhliang@zjut.edu.cn}

\renewcommand{\shortauthors}{Zhang et al.}

\begin{abstract}
As programming education becomes more widespread, many college students from non-computer science backgrounds begin learning programming. Collaborative programming emerges as an effective method for instructors to support novice students in developing coding and teamwork abilities. However, due to limited class time and attention, instructors face challenges in monitoring and evaluating the progress and performance of groups or individuals. To address this issue, we collect multimodal data from real-world settings and develop \textit{CPVis}, \RR{an interactive visual analytics system designed to assess student collaboration dynamically.} Specifically, \textit{CPVis} enables instructors to evaluate both group and individual performance efficiently. \textit{CPVis} employs a novel flower-based visual encoding to represent performance and provides time-based views to capture the evolution of collaborative behaviors. A within-subject experiment (N=22), \RR{comparing \textit{CPVis} with two baseline systems, reveals that users gain more insights, find the visualization more intuitive, and report increased confidence in their assessments of collaboration.}
\end{abstract}





\maketitle

\section{Introduction} 
\label{sec:introduction}
Computing and programming have become integral components of the foundational curricula in many countries, spurring the rapid expansion of introductory programming courses, \RR{which are frequently taught in large-scale classrooms with hundreds of students~\cite{zhang2024cflow,hawlitschek2023empirical}.
To manage these large classes,} instructors employ diverse teaching methods, such as collaborative learning and flipped classrooms, to coordinate better and manage students~\cite{yan2024evidence}.  
In computer-supported collaborative learning (CSCL), \RR{students collaborate in groups, engaging in communication and interactions to perform high-level cognitive tasks}, such as learning programming languages, solving programming problems, and enhancing logical reasoning and critical thinking skills~\cite{ouyang2022applying}.

\RR{
Although research highlights the benefits of CSCL\cite{ouyang2022applying}, instructors still face challenges in practice. In large collaborative programming classes, they must balance guiding multiple groups with fostering student engagement. This makes it difficult to closely monitor group dynamics and individual performance, and to offer real-time feedback during discussions\cite{mangaroska2022exploring}. Additionally, assessments often focus on final solutions rather than the collaborative programming process~\cite{schulz2023exploring,yan2019pensieve}, limiting instructors' ability to provide feedback on techniques or evaluate individual engagement, thus complicating the assessment of collaborative programming~\cite{schulz2023exploring}.
}
Most current research focuses on evaluating general group discussions~\cite{niemantsverdriet2017recurring}, analyzing multi-person dialogue to extract discussion topics and participant behavior patterns. These studies propose tools such as context-based narrative meeting dashboards~\cite{samrose2021meetingcoach} and visual discourse analysis reports~\cite{sevastjanova2021visinreport}, significantly improving the efficiency of group discussion evaluation. However, evaluating collaborative programming in large classrooms requires more comprehensive and multi-level analysis.

Collaborative programming is a dynamic process involving the gradual evolution of problem-solving, meaning-making, and knowledge construction~\cite{ma2022detecting}. 
\RR{Within collaborative programming groups, students often switch between roles, such as the ``Driver,'' who writes the code; the ``Navigator,'' who provides ideas; and the ``Monitor,'' who checks each line of code for issues~\cite{lewis2011pair}. These role dynamics and group interactions significantly impact students' engagement.}
Moreover, effective evaluation requires considering not only students' role changes but also group-level dynamics and collaboration. Therefore, studying students' engagement, behavior patterns, and collaborative problem-solving skills from a process-oriented perspective is essential for understanding the complexities of collaborative programming.
\RR{To address this pressing need, we aim to develop a visual analysis system that harnesses multimodal data and advanced visualization techniques, empowering instructors with comprehensive insights into collaborative programming and enabling them to evaluate group and individual student performance with greater efficiency.}

However, developing such a system involves overcoming three key challenges. 
First, extracting potential problem-solving patterns from unstructured data and analyzing their dynamic changes over time is challenging. In addition to semantic analysis of speech, \RR{raw data such as screen recordings and behavioral videos in programming education complicate the analysis, while students' diverse and constantly changing roles in collaborative processes make it difficult to capture and identify role transitions accurately.
Second, the visual design must balance intuitive representation and complex data encoding (e.g., multidimensional, temporal, and comparative) while presenting individual students and groups cohesively rather than in isolation.}  
Third, instructors need to combine individual student changes with group patterns to gain deep insights into the dynamics of collaborative problem-solving.
\RR{To address the first challenge, we leverage large language models (LLMs) for fine-grained semantic analysis and error detection of unstructured data, enabling the extraction of students' dynamically changing roles while reducing the time required for manual annotation~\cite{hou2024codetailor}. }
Additionally, we construct a learning analytics framework to evaluate the performance of groups and individuals in collaborative programming.  
For the second challenge, we design novel flower-based visual metaphors to present student and group performance intuitively. We also use timeline visualizations to reveal dynamic changes in key features and patterns.  
For the third challenge, we introduce \textit{CPVis}, an interactive visual analytics system that provides instructors with perspectives ranging from overviews to detailed insights, enabling the exploration of multidimensional learning dynamics in collaborative programming. 

The contributions of this study are summarized as follows:  

\begin{itemize}
	\item We propose an innovative set of visual metaphors to create profiles for students and groups, facilitating the evaluation and comparison of different groups and individual students in collaborative programming.

	\item We develop \textit{CPVis}, an interactive visual analytics system designed for instructors to assess students' problem-solving processes and collaborative programming dynamics, \RR{facilitating quantitative assessment through the use of LLMs }and supporting the exploration of collaborative programming performance from the holistic (group level) to the detailed (individual level).

	\item We conduct quantitative, case, and user studies, demonstrating that \textit{CPVis}' multimodal learning analytics are intuitive and effective, enabling instructors to quickly and efficiently evaluate collaborative programming.

\end{itemize}

\section{Related Work}
\label{sec:related}
\RR{In this section, we discuss the relevant research, including Programming  Education and Evaluation, Multimodal Learning Analysis in Collaborative Programming, and Visual Analysis of Collaborative Behaviors in Meeting.}

\subsection{Programming Education and Evaluation}
As technology advances rapidly, programming skills are a significant driving force behind social and economic development. 
Whether in artificial intelligence~\cite{breideband2023community}, big data~\cite{rasul2023role}, or the Internet of Things~\cite{deline2021glinda}, programming plays a crucial role in these cutting-edge technologies.
The importance of programming education increases globally, with programming being integrated into the foundational curricula of many countries~\cite{yan2019pensieve}. 
Instructors supplement programming education by incorporating lab courses that encourage coding practice~\cite{park2018elicast}. \RR{Among the various teaching strategies employed in these lab courses, collaborative programming emerges as a commonly used and effective method for teaching novices, fostering teamwork and practical coding skills~\cite{gaudencio2014can}.}

Though programming education is widespread, evaluating programming skills in a large classroom remains challenging, particularly in assessing students' code~\cite{zhang2024cflow}. 
Researchers develop tools that leverage LLMs to help instructors analyze code more effectively~\cite{hellas2023exploring,leinonen2023comparing}. 
These tools provide insights into students' code~\cite{glassman2015overcode,guo2015codeopticon} and facilitate code to identify patterns~\cite{mathew2020slacc,mathew2021cross,zhang2023runex}. 
For example, VizProg uses CodeBERT to cluster solutions and visualize coding progress in real-time~\cite{zhang2023vizprog}, while CFlow combines LLMs for semantic annotation and interactive views to identify patterns in courses~\cite{zhang2024cflow}. 
\RR{However, existing tools largely assessed code fragments from collaborative programming in isolation, failing to consider the collaborative behaviors and dynamics among students~\cite{tang2024vizgroup}.}
In contrast, \textit{CPVis} analyzes final code submissions and students' collaborative behaviors, enabling instructors to provide detailed, evidence-based feedback that enhances students' metacognitive awareness and addresses deficiencies in their learning approaches.

\subsection{Multimodal Learning Analysis in Collaborative Programming}
In collaborative programming, students work in groups to complete programming tasks, share code modifications, and review each other's work~\cite{maguire2014enhancing,hou2024automated}. 
This method aims to improve students' programming efficiency, code quality, and teamwork skills through collaboration~\cite{hanks2011pair,diana2017instructor}. 
To understand students' interactions and engagement, multimodal data such as conversations and screen recordings from the collaborative programming process are typically collected~\cite{mangaroska2022exploring,wang2024unveiling}.
Multimodal Learning Analytics (MMLA) provides new insights into students' learning by analyzing various data streams generated during the learning process, such as speech, facial expressions, and gestures~\cite{yan2024evidence,chejara2024bringing}. 
For instance, multimodal analysis in collaborative learning is pioneered by quantifying learners' nonverbal behaviors through video segmentation to predict academic performance~\cite{ambady1993half}.
Various types of multimodal data are increasingly used to analyze collaborative problem-solving processes in learning environments~\cite{zhang2022towards,ma2022detecting}. 
\RR{For example, Mangaroska et al.~\cite{mangaroska2022exploring} combine multimodal data (camera, electroencephalogram, eye-tracking) with cognitive load theory and affective dynamics models to analyze problem-solving in collaborative learning.}
Although these studies explored the analysis of problem-solving processes, they are limited to extracting relevant metrics from multimodal data, \RR{lacking the capability to dynamically analyze students' problem-solving processes over time in a narrative form. }

Collaborative programming is a complex form of collaborative learning, involving various factors such as role transitions during programming~\cite{lewis2011pair} and interaction patterns~\cite{zhou2024understanding}. 
\RR{For instance, Lewis et al. accelerate the speed at which students complete exercises by intervening in the roles students play in collaborative programming~\cite{lewis2011pair}.
However, existing research fails to address the interdependent development of other key dimensions such as teacher scaffolding~\cite{ouyang2022applying} and learning engagement analysis~\cite{hou2024codetailor}\cite{wu2019analysing}, focusing instead on the impact of a single dimension on the collaborative programming process~\cite{rosen2020towards}.}
These limitations highlight the need for designing an innovative visualization framework to address the unique challenges of tracking multiple metrics over time, clearly representing interaction patterns in collaborative programming to help instructors better understand and evaluate the collaborative problem-solving process.

\subsection{Visual Analysis of Collaborative Behaviors in Meeting}
Collaborative programming discussion is a form of multi-person collaboration, and we focus on analyzing collaborative behavior during meetings.
In the visualization community, collaborative behavior analysis is explored across various applications, including face-to-face~\cite{bachour2010interactive} and online discussions~\cite{wang2024meeting}, as well as real-time~\cite{son2023okay} and post-meeting analysis~\cite{shi2018meetingvis}. 
\RR{These two dimensions can be combined to form a larger design space.} For instance, integrating face-to-face discussions with real-time analysis reduces environmental distractions and fosters more natural, seamless interactions~\cite{aseniero2020meetcues}.

Some real-time analysis systems~\cite{chen2023meetscript,chandrasegaran2019talktraces,shi2017ideawall}, such as ClassBeacons, effectively visualize instructors' time and attention allocation through light objects on students' desks, streamlining issue resolution during student discussions~\cite{an2019classbeacons,an2018classbeacons}. 
Real-time analysis systems focus on summarizing ongoing discussions without disrupting their flow~\cite{haliburton2023walking}. Similarly, Groupnamics provides an overview of parallel groups in online classrooms, helping identify those needing intervention~\cite{sato2023groupnamics}.
However, post-analysis offers a distinct advantage for task evaluation, as it provides a broader perspective to understand better and assess the outcomes of collaboration.
Some post-analysis visualization systems, such as ConToVi, NEREx, and Meeting Mediator~\cite{el2016contovi,el2017nerex,kim2008meeting}, focus on speaker behavior patterns over time but do not fully capture the evolution of speakers' problem-solving abilities across different discussion topics. \RR{Moreover, these systems primarily summarize single-session dialogues, lacking the ability to analyze multiple groups or discussions simultaneously.}

Studies analyzing group dialogues in classrooms typically focus on engagement and comprehension but often lack detailed visual insights into the underlying problem-solving logic, strategies, or intentions~\cite{fujii2018sync,macneil2019ineqdetect,tausch2014groupgarden}.
\RR{In this work, we focus on the post-analysis of face-to-face discussions, emphasizing the collaborative problem-solving process across multiple groups. 
\textit{CPVis} incorporates narrative-based visualizations to depict and compare the evolution of dialogue, behavior, and engagement, providing instructors with richer evidence to assess group dynamics and individual engagement more effectively.}

\section{Formative Study} 
\label{sec:formative}
We conducted a formative study to explore instructors' challenges in collaborative programming and their visualization needs, with the study protocol approved by our university's Institutional Review Board (IRB).
 
 \subsection{Participants}
We recruited ten participants (five females, age: \(30.6 \pm 6.6\)) with collaborative programming experience, divided into three groups: two educational technology experts (E1, E2) and eight instructors (T1–T8) with an average of 8.14 years of teaching programming. Participants, recruited via snowball sampling from the authors' network, received \$20 as compensation.

\subsection{Procedure}\
\RR{The formative study used semi-structured interviews conducted via Zoom, divided into two sections: (A) Questions and Answers and (B) Ratings.}
\textit{A: Questions and Answers.} 
Each participant was independently interviewed to explore the need for collaborative programming analysis. Topics included teaching experience with collaborative programming, class organization, challenges faced, assessment methods, and group and individual performance evaluations. Follow-up questions were asked for clarification or more profound insights. Each session lasted 40–60 minutes and was documented through written notes, audio, and video recordings.
\textit{B: Rating.} 
Participants then rated two aspects of the collaborative programming visual analytics system. They rated feature importance (Q1) on a 1–7 scale (7 being the highest) and ranked features by priority (Q2). 
\RR{More details are in appendix A.}

\subsection{Findings} 

\begin{figure*}[htbp]
	\centering
	\includegraphics[width=1\linewidth]{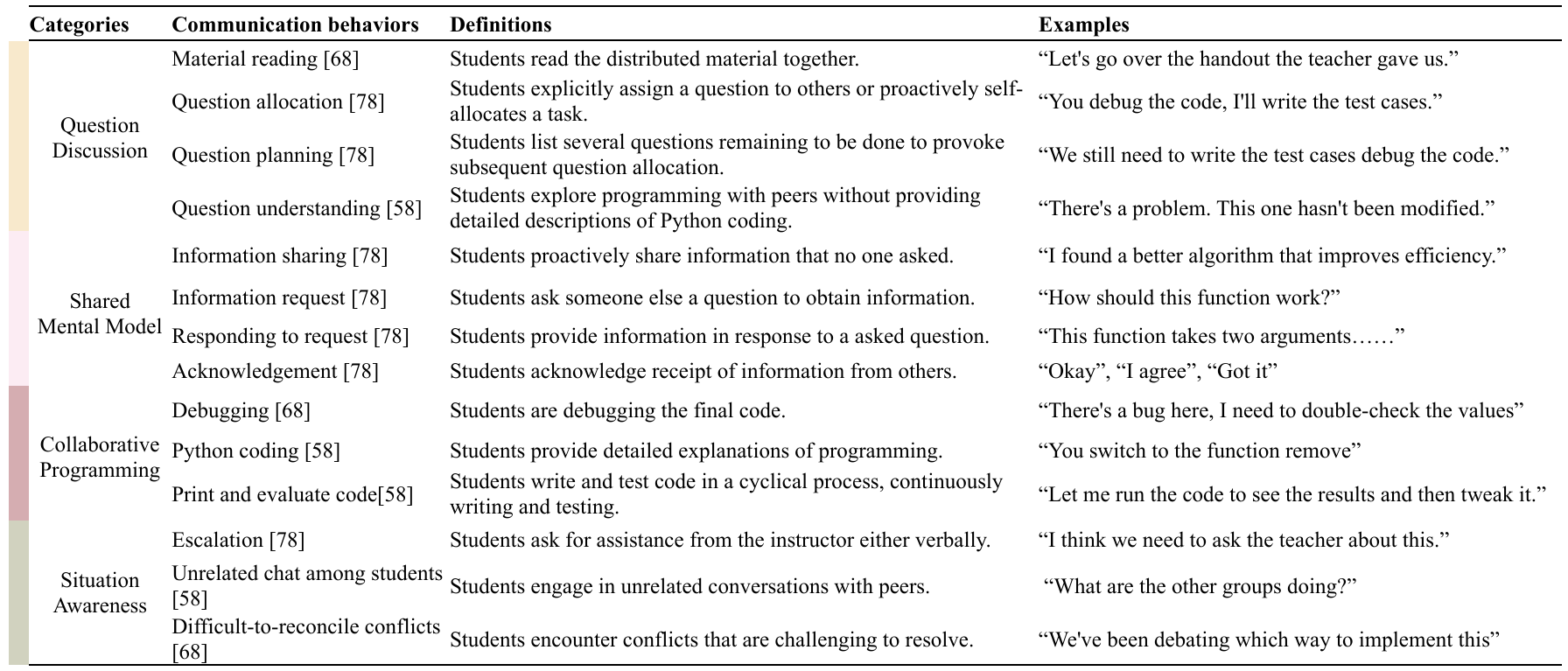}
	\caption{Collaborative programming coding schemes, along with their definitions and examples.}
        \Description{Collaborative programming coding schemes, along with their definitions and examples.}
	\label{fig:scheme}
\end{figure*}

\subsubsection{Evaluation of Groups:}
In large classes, instructors struggled to monitor each group without the help of teaching assistants. They often relied on group presentations (E2, T1, T5), but it was inaccurate. Limited class time forced instructors to focus on solving issues rather than actively monitoring groups.

\textbf{Group Performance:}
Evaluating group work was based on the final code rather than the process, limiting the ability to provide feedback (T7). Some instructors used presentations or technical documentation to streamline assessments (T6). However, evaluating more than 20 groups was still challenging, and instructor assistance was often overlooked during evaluations.

\textbf{Collaboration Evaluation:}
Effective collaboration wasn't just about task completion speed and group dynamics. Some groups (T7) completed tasks quickly due to one member's efforts, not true collaboration. 
\RR{E1 proposed the collaborative problem-solving framework to distinguish task effectiveness from team effectiveness.} Instructors (T6) believed monitoring discussion time helped assess collaboration quality and task difficulty, but off-topic discussions made it hard to evaluate group discussions (E1, T3, T5).

\subsubsection{Evaluation of Students:}

In large classes, assessing individual contributions in group work was challenging. Instructors relied on peer evaluations (T2, T3, T4) and self-reported task distributions (E2), which were often subjective.

\textbf{Individual Performance:}
Instructors typically reviewed code to assess understanding, but measuring individual contributions in group work was hard. Leadership roles often reflected a deeper grasp of concepts (T1), but tracking individual engagement was difficult in large classes.

\textbf{Personalized Feedback:}
Providing personalized feedback was difficult, as group results often masked individual struggles. T3 and T4 noted that group collaboration fostered peer learning but could lead to less engagement from weaker students.
\textit{T8 added that offering personalized feedback in large classes was time-consuming and burdensome.}

\subsection{Design requirement}
\label{dr}
Based on the interview findings, we identified six design requirements (R1–R6) across three levels, summarized as follows:

Support \textbf{\textit{inter-group-level}} to provide a macro perspective, enabling instructors to observe the overall situation of all groups comprehensively and fully understand class-wide dynamics.

\textbf{R1: Displaying the Overall Performance of all Groups.} 
Instructors face challenges in supervising multiple groups simultaneously and shifting focus efficiently. Participants stressed the need for an overview of group performance, allowing instructors to grasp class dynamics and selectively review specific groups.

\RR{\textbf{R2: Comparing Similar and Different Groups.} 
Instructors often compare students' performance to assess their relative standing within the class~\cite{marsh1997making}. Such comparisons enable a more accurate evaluation of group performance and help identify groups excelling or encountering challenges.}

Supporting \textbf{\textit{intra-group-level}} visual exploration to offer a meso perspective, enabling instructors to observe specific groups' performance and gain a comprehensive understanding of group dynamics during the collaborative programming process.

\textbf{ R3: Understand the Dynamics of Programming Problem Solving.}
Analyzing a group's evolving communication patterns and computational thinking during programming tasks provides instructors with deeper insights into students' progress and intermediate learning objectives—details missed in final submissions alone.

\textbf{ R4: Identify Teacher Scaffolding in Collaboration.}
Instructors play a vital role in guiding groups during collaborative programming. Understanding the scaffolding provided and students' responses can help refine instructional strategies, improving the overall effectiveness of collaborative programming.

Supporting \textbf{\textit{individual-student-level}} visual exploration to provide a micro perspective allows instructors to observe each student's performance within a specific group and better understand their role and collaboration.

\textbf{R5: Track Changes in Student Engagement Over Time.} 
Limited classroom time makes it challenging for instructors to monitor individual student engagement in programming tasks.
Tracking and visualizing engagement trends is essential for assessing performance and refining instructional practices.

\textbf{R6: Access Detailed Raw Data.} 
Instructors require access to raw data, such as collaboration videos, conversations, and background information. These details are crucial for validating analysis results and supporting personalized feedback and assessments.

\begin{figure*}
    \centering
    \includegraphics[width=1\linewidth]{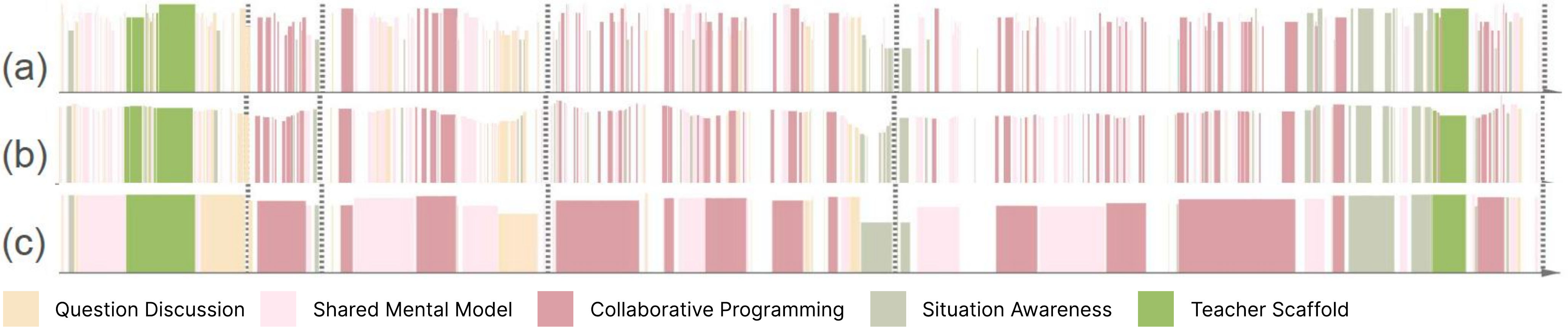}
    \caption{(a) shows the bar chart of the raw data, (b) presents the results of applying Moving Average Smoothing to reduce anomalies in prediction percentages, and (c) highlights the reduction of visual clutter and emphasizes sequential behavior patterns after merging behaviors of the same category.}
    \label{fig:bar}
    \Description{(a) shows the bar chart of the raw data, (b) presents the results of applying Moving Average Smoothing to reduce anomalies in prediction percentages, and (c) highlights the reduction of visual clutter and emphasizes sequential behavior patterns after merging behaviors of the same category.}
\end{figure*}

\section{Data Collection and Processing}
\label{sec:data}
\RR{In this section, we provided an overview of the data collection context and introduced the collaborative programming performance framework along with its metric quantification methods.}

\subsection{Data Collection}
We collaborated with Professor E1, an expert in programming education, and teaching assistants (TA1 and TA2), experienced in Python, to collect data from E1's Spring 2023 Python course with 66 non-computer science freshmen in 22 groups. Using non-intrusive methods, we recorded group discussions, screen activities (without audio), and code submissions. Session lengths ranged from 10 to 60 minutes based on question completion. 
Due to data quality issues, we selected data from 19 groups (57 students) for analysis.

\subsection{Data Preprocessing}
In collaborative programming analysis, students' spoken content was key to understanding discussion and evaluating collaboration. We used the Faster-Whisper model~\cite{fasterwhisper} for speech recognition and the Pyannote-audio model~\cite{pyannoteaudio} for speaker diarization. 
For groups lacking clear problem-solving strategies, we used Tesseract OCR~\cite{tesseract} to analyze screen recordings and extract key frames through screenshots.

\subsection{Scope of Collaborative Programming Performance Framework}
Evaluating student and group performance in collaborative programming required considering multiple dimensions~\cite{hawlitschek2023empirical}.  
Building on literature and expert input (E1), we proposed the following comprehensive analytical framework to assess performance.

\subsubsection{Student Performance Assessment}
\label{shema}
Previous research demonstrated that students' skills, backgrounds, and personalities in the classroom vary significantly, affecting their engagement and learning outcomes~\cite{wu2019analysing}. 
Therefore, we focus on each student's \textit{background} (prior academic performance and major), \textit{role transitions}, \textit{behavioral engagement}, and \textit{cognitive engagement}.

\textbf{Problem-solving Categorization:}
Based on previous frameworks~\cite{wu2019analysing}, team theory~\cite{zhao2023analysing}, and collaborative coding processes~\cite{sun2021three}, we developed a coding scheme (Fig.~\ref{fig:scheme}) to capture group problem-solving in collaborative programming. 
The scheme used four color-coded categories to represent discussion types. 
The first three categories followed a hierarchical structure, indicating discussion depth, while the green category focuses on situation awareness and specific behaviors.

Building on the scheme, we used tailored prompts with the ChatGPT-4o model~\cite{gpt4o} to classify behavioral patterns in transcribed dialogue \RR{(More details are in appendix B)}. 
\RR{The model provided a prediction percentage of uncertainty for each classification, improving result interpretability. }
To minimize anomalies, we applied a ``moving window'' technique with Moving Average Smoothing~\cite{chang2022muse}, stabilizing prediction percentages (Fig.\ref{fig:bar}-b). To reduce visual clutter in long time-series data, we aggregated consecutive instances of the same category, averaging prediction percentages (Fig.\ref{fig:bar}-c). These results were displayed in the timeline panel's progress bar, enabling detailed analysis by zooming into specific behavior categories in Sec.~\ref{barchart}.

\textbf{Roles Extraction:}
We analyzed each speaker's dynamic roles (Driver, Navigator, and Monitor) during programming~\cite{lewis2011pair}. Using ChatGPT-4o and prompts based on the Thought Chain Model~\cite{wei2022chain}, we guided the model through step-by-step reasoning to generate role classifications. Prompts were iterated for clarity, and the model's responses were structured hierarchically and returned in JSON format. Each query was repeated ten times, with the majority result adopted for classification.

\RR{\textbf{Behavioral Engagement:} reflected the level of effort and participation students invested in learning~\cite{fredricks2022measurement}. 
In our study, we focused on the duration and frequency of student speech.} 
We extracted conversation data, excluding irrelevant chat, and divided each conversation into two parts: the first half and the full conversation. We then measured speaking duration, frequency, and degree centrality using co-occurrence networks~\cite{ng1999toward}. For each question, we created and normalized two networks, followed by Non-negative Matrix Factorization (NMF)~\cite{lee2000algorithms} to identify key behavioral patterns for dynamic group comparison.

\RR{\textbf{Cognitive Engagement:} referred to the cognitive investment students made in their learning. We highlighted the role changes and behavior frequencies of students during the collaborative process. }
To capture dynamic changes in student cognitive engagement, we split the dialogue for each question into two segments: the first half and the full dialogue. We extracted the frequency of each speaker's 14 behavioral categories and their roles at each timestamp. After normalizing these features for consistency, we applied NMF to reduce dimensionality and assess each speaker's cognitive engagement.

\begin{figure*}
  \includegraphics[width=\textwidth]{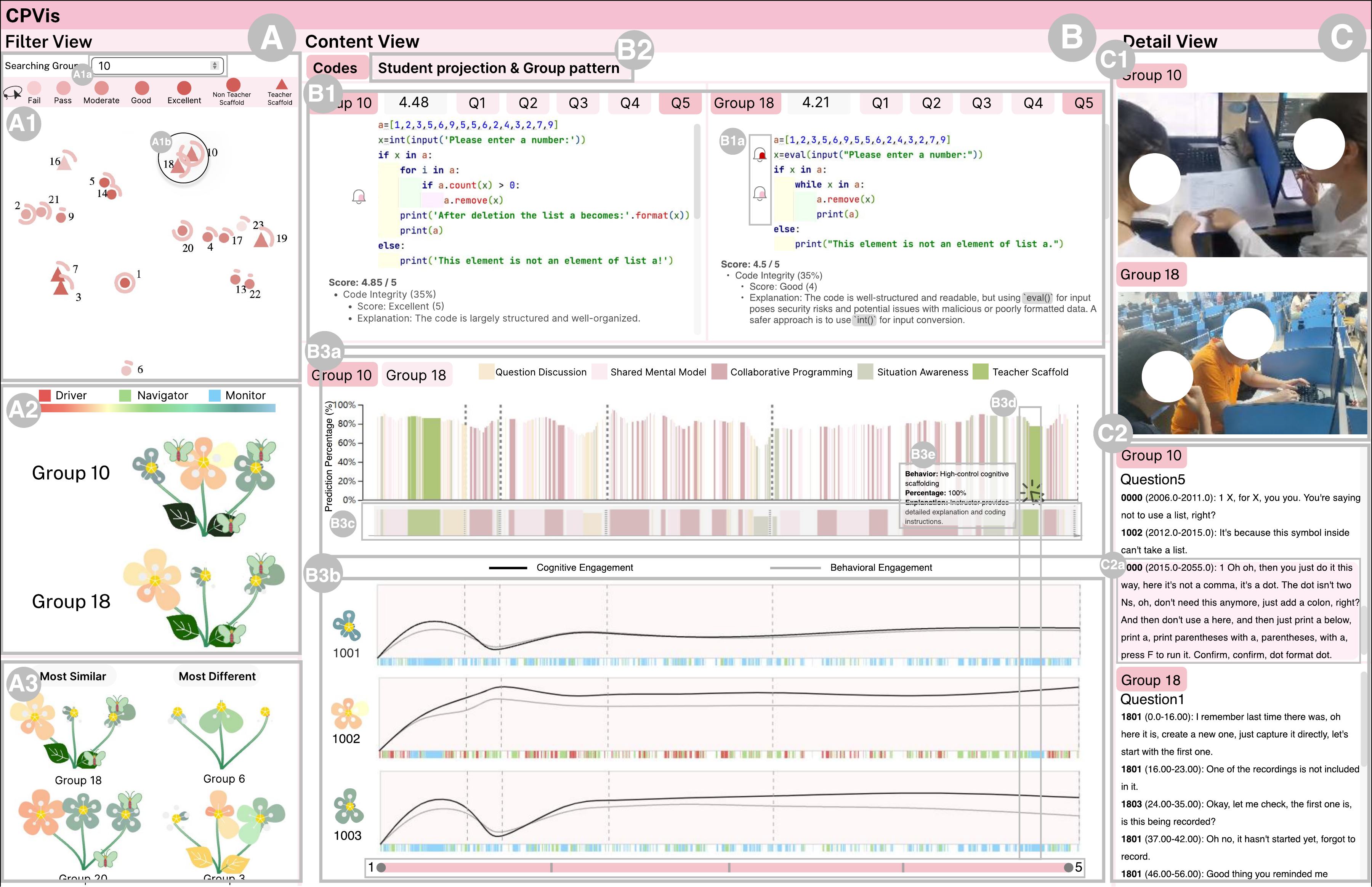}
  \caption{\RR{A screenshot of Group 10 view.} \textit{CPVis} applies multimodal learning analysis to provide instructors with evidence for evaluating group and student performance. It consists of three views:
Filter View (A) Provides an overview and allows group selection. The selected groups appear in the lasso selection area (A2), and the similarity panel (A3) displays the most similar and different groups based on the search (A1a).
Content View (B) Displays group performance, with the B1 panel showing completed codes, the B3a panel illustrating the behavior sequence, and the B3b panel showing student engagement over time.
Detail View (C) Presents the group's collaborative programming video (C1) and raw conversation data (C2).}
  \Description{A screenshot of Group 10 view. \textit{CPVis} applies multimodal learning analysis to provide instructors with evidence for evaluating group and student performance. It consists of three views:
Filter View (A) Provides an overview and allows group selection. The selected groups appear in the lasso selection area (A2), and the similarity panel (A3) displays the most similar and different groups based on the search (A1a).
Content View (B) Displays group performance, with the B1 panel showing completed codes, the B3a panel illustrating the behavior sequence, and the B3b panel showing student engagement over time.
Detail View (C) Presents the group's collaborative programming video (C1) and raw conversation data (C2).}
  \label{fig:teaser}
  \end{figure*}

\subsubsection{Group Performance Assessment}
We evaluated group performance based on three dimensions: code quality, collaborative problem-solving, and teacher scaffolding. 
Through in-depth discussions with domain experts, we assessed how each dimension was valued and measured in the context of our study.

\label{code}
\textbf{Code quality}, reflecting students' mastery of course concepts, was a key metric for evaluating group performance. To assess student submissions, we used ChatGPT-4o~\cite{gpt4o} to evaluate dimensions such as problem-solving, code integrity, accuracy, and algorithmic innovation, scoring each on a 1–5 scale. After refining evaluation prompts, we ran the assessment ten times per submission, averaging the results to ensure consistency and reliability.

\textbf{Collaborative Problem-Solving (CPS):} 
Earlier studies categorized CPS into team effectiveness and task effectiveness~\cite{rosen2020towards}. Team effectiveness was measured by student engagement, while task effectiveness was assessed through code quality. 
To evaluate CPS, we examined task effectiveness, represented by the average question score (\(\bar{s}\)), and team effectiveness, assessed through the standard deviation of engagement (\(\sigma_e\)) and the average engagement score (\(\bar{e}\)) as shown in Equation \ref{eq:1}. We then used the coefficient of variation (\(CV_e\)) \RR{to account for both engagement variability and engagement}. Finally, the overall collaboration quality was calculated using Equation \ref{eq:2}, combining question performance and engagement balance. 
\begin{equation}
\sigma_e = \sqrt{\frac{1}{n} \sum_{i=1}^{n} (e_i - \bar{e})^2}, \quad CV_e = \frac{\sigma_e}{\bar{e}}
\label{eq:1}
\end{equation}

\begin{equation}
Quality = \bar{s} \cdot (1 - CV_e)
\label{eq:2}
\end{equation}
As shown in Table \ref{table:comparison}, Group 19, despite achieving a respectable average score, exhibited imbalanced engagement, leading to a lower collaboration quality score. In contrast, Group 20 demonstrated more balanced and higher engagement, resulting in a better overall collaboration quality.
\begin{table}[htbp]
\centering
\begin{tabular}{cccccc}
\toprule
\textbf{Group} & \(\bar{s}\) & \textbf{Engagement Levels} & \(\sigma_e\) & \(\text{CV}_e\) & \textbf{CQ} \\
\midrule
Group 19 & \(4.11\) & (10.515, 9.725, 4.575) & \(2.80\) & \(0.24\) & \(2.80\) \\
Group 20 & \(4.14\) & (10.06, 9.32, 8.62) & \(0.73\) & \(0.08\) & \(3.88\) \\
\bottomrule
\end{tabular}
\caption{Comparison of Group 19 and Group 20 on Collaboration Quality (CQ).}
\label{table:comparison}
\end{table}

\textbf{Teacher Scaffolding,} categorized into cognitive (low, medium, high-control) and metacognitive forms~\cite{ouyang2022applying}, reflected the level of support provided to a group and its impact on programming performance. We evaluated four scaffolding dimensions, leveraging GPT-4o for annotation. By using targeted prompts and examples, we improved classification accuracy, while teacher scaffolding was categorized according to the type of support based on a semantic analysis of interactions.

\section{System Design} 
\label{sec:vis}

In this session, we introduce \textit{CPVis} (Fig.~\ref{fig:teaser}), a web-based visual analytics system to assist instructors in evaluating collaborative programming.
\subsection{System Overview}
\textit{CPVis} is a comprehensive system supporting multi-level, progressive analysis, from group-level interactions to individual student performance. Instructors can select specific groups for focused analysis, such as comparative evaluations (R2). Additionally, \textit{CPVis} offers a drill-down feature, enabling an overview of student collaboration and detailed insights into individual performance.
\textit{CPVis} includes four main components:
\textbf{Initial Selection}, instructors can select groups using the lasso tool or search function in the ``Group Overview'' view (Fig.\ref{fig:teaser}-A1). The system displays an overview of the selected groups (Fig.\ref{fig:teaser}-A2), compares similar and different groups (Fig.\ref{fig:teaser}-A3), and synchronizes updates across views (Fig.\ref{fig:teaser}-B).
\textbf{Drill-Down Analysis}, instructors can examine the group's code (Fig.\ref{fig:teaser}-B1) and analyze problem-solving approaches in the Content View. The interaction pattern panel (Fig.\ref{fig:student}-B2b) reveals behavior patterns, while the timeline panel (Fig.\ref{fig:teaser}-B3a) shows activity sequences. The student overview panel (Fig.\ref{fig:student}-B2a) compares individual performance across the class, and the timeline panel (Fig.~\ref{fig:teaser}-B3) highlights engagement and role changes.
\textbf{Multi-Level Interaction}, the system's layered visualization allows instructors to explore and analyze both group and individual behavior, enabling precise assessment of the collaboration process. It also supports side-by-side group comparisons.
\textbf{Detailed Review}, the detail view (Fig.~\ref{fig:teaser}-C) provides original discussion videos and transcripts, enabling in-depth analysis of student conversations and problem-solving processes.
\begin{figure}
	\centering
	\includegraphics[width=1\linewidth]{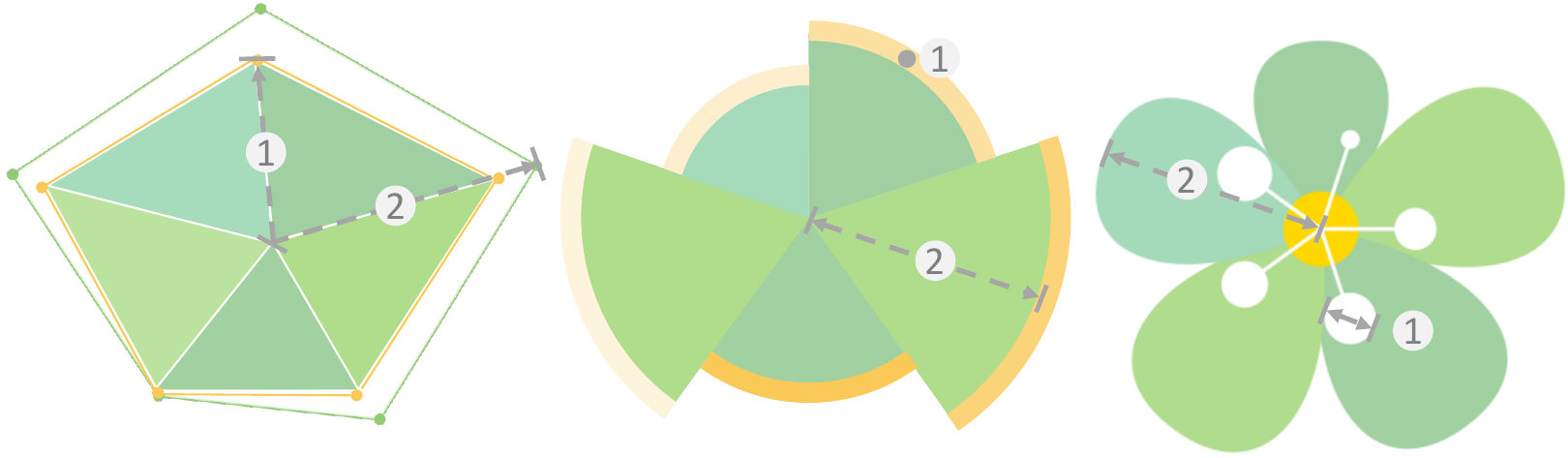}
	\caption{The iterative design of student glyphs: 1 represents cognitive engagement, and 2 represents behavioral engagement.}
        \Description{The iterative design of student glyphs: 1 represents cognitive engagement, and 2 represents behavioral engagement.}
	\label{fig:designl}
\end{figure}
\subsection{Visual Design}
\begin{figure*}
	\centering
	\includegraphics[width=1\linewidth]{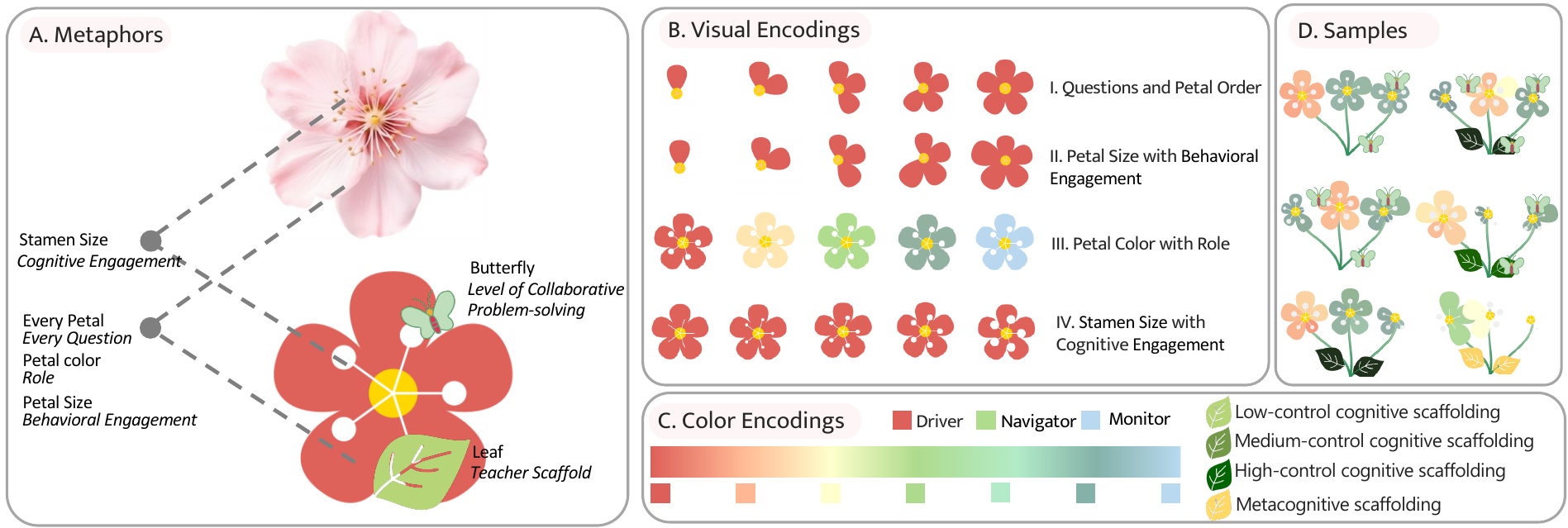}
	\caption{The flower metaphor in \textit{CPVis}, along with its visual encoding, color coding, and some samples.}
        \Description{The flower metaphor in \textit{CPVis}, along with its visual encoding, color coding, and some samples.}
	\label{fig:encoding1}
\end{figure*}

We iterated continuously during the glyph design process to optimize the visual representation. Initially, we used a star-shaped design (e.g., radar charts), as shown in Fig.~\ref{fig:designl}. While radar charts effectively displayed behavioral and cognitive engagement, dividing the chart into sections for individual tasks introduced an unnecessary dimension (shape size) that was meaningless and prone to misinterpretation. 
We then shifted to a circular design, encoding behavioral engagement as the radius of a sector and cognitive engagement as the color of the outer ring. However, this design had a significant flaw: color mapping was less intuitive than size mapping, and using color saturation as a visual channel lacked precision. Additionally, both designs struggled to combine individual students into group glyphs intuitively.
\RR{Inspired by previous research~\cite{tausch2014groupgarden,tausch2016comparison,xiong1999peoplegarden}, we introduced a visual design inspired by the flower metaphor, where students are represented as flowers (Fig.~\ref{fig:encoding1}).
The size of the petals represented behavioral engagement, while the size of the stamen indicated cognitive engagement. Three colors were used to represent three different roles, and the varying colors of the leaves symbolized different levels of teacher scaffolding. The number of butterflies reflected the level of collaborative problem-solving ability. As a result, the overall group glyph naturally took the form of a bouquet (R4).}
This approach resolved the issue of merging individual students into group glyphs while enhancing the design's readability and intuitiveness (R1). 
\RR{The final design struck a balance between aesthetics and functionality, effectively conveying the performance of individual students and groups, allowing users to quickly compare similar or different groups (R2).}

\subsection{Filter View}
The Filter View (Fig.~\ref{fig:teaser}-A) serves as the starting point for analysis, featuring an interactive projection panel (Fig.~\ref{fig:teaser}-A1) and a similarity panel (Fig.~\ref{fig:teaser}-A3) to help users filter and explore groups of interest. 
The projection panel displays the distribution of groups in a 2D space to reveal clustering patterns and outliers. We apply the t-SNE algorithm to maximize separation between dissimilar groups, creating clearer clusters. 
To avoid visual clutter caused by group glyphs in dimensionality-reduced views, we follow the approach of Tac-Miner~\cite{wang2021tac}, representing groups as points or rectangles based on whether they received teacher scaffolding, with color coding reflecting prior performance. 
\RR{Additionally, the outer arc represents the duration of the discussion.}
Users can select groups using the lasso tool (Fig.~\ref{fig:teaser}-A1b) or search for specific groups (Fig.~\ref{fig:teaser}-A1a). The similarity panel (Fig.~\ref{fig:teaser}-A2) shows the most similar and dissimilar groups based on Euclidean distance.

\subsection{Content View}
Once the search group is selected, users can perform a detailed analysis through the content view (Fig.~\ref{fig:teaser}-B). This view comprises four panels, allowing for a layered exploration of group and individual student performance. 

\subsubsection{Codes Panel}
In the upper left corner of the content view, a control button (Fig.~\ref{fig:teaser}-B1) allows users to toggle between the Codes panel and the Student Projection/Group Pattern panels. Below, users can compare the code quality between the selected group (left) and a comparison group (right). 
For instance, Figure~\ref{fig:teaser}-B1 shows Group 10's answer to Question 5 (left) and Group 18's answer (right). Hint boxes provide two types of feedback: red tips indicating code deficiencies (Fig.~\ref{fig:teaser}-B1a) and pink tips showing that a method was learned from pre-class materials, signaling comprehension, and application. 
Below the code, the ChatGPT-4o score and its rationale are provided, enabling quick, in-depth code evaluation and highlighting areas where groups faced challenges.

\subsubsection{Students Projection (Fig.~\ref{fig:student}-B2a)}

We project students from different groups using t-SNE for clustering, highlighting students with similar performance. 
The three flowers representing the search group are connected with dashed lines to clarify group member distribution and similarity, helping users assess group homogeneity or heterogeneity. 
Only the comparing group's flowers are highlighted to minimize visual clutter. 
\RR{Users can zoom in/out to explore specific students and view detailed background information (e.g., major, grades) by hovering over individual points (Fig.~\ref{fig:student}-B2c).}

\subsubsection{Group Pattern Panel (Fig.~\ref{fig:student}-B2b \& d)}
We use Epistemic Network Analysis (ENA)~\cite{zhao2023analysing} to analyze the dynamic connections between cognitive elements and the collaborative problem-solving behaviors of groups (R3). 
In the Group Pattern Panel, each node represents a behavior in the collaborative problem-solving process, with colors following the coding scheme in Section~\ref{fig:scheme}. Node size indicates behavior frequency, and the intensity of the color reflects the frequency of interactions between behaviors. More significantly, darker nodes represent more frequent behaviors and interactions. 
Users can click on different question buttons to examine dynamic changes in behavior across specific questions or select multiple questions to observe how behaviors evolve during transitions between tasks. Hovering over nodes reveals detailed information about each behavior.
\RR{When comparing two groups, the system displays side-by-side behavior networks.} 
The Group Pattern Panel displays the search group's behavior patterns and compares them to those of the comparison group during the collaborative problem-solving process (Fig.~\ref{fig:student}-B1b).

\subsubsection{Timeline Panel}
\label{barchart}
Users can toggle between the search and comparing groups using the control button in the upper left corner of the Timeline Panel (R5). This panel displays group and individual student performance over time using a filterable bar chart and line chart (Fig.~\ref{fig:teaser}-B3).
In the bar chart, each bar represents a timestamp, with colors indicating different collaborative problem-solving behaviors. The bar height reflects uncertainty, as calculated by ChatGPT-4o. Dashed lines separate different questions for visual clarity. Hovering over a bar reveals the behavior category, predicted certainty, and reasoning behind the label (Fig.~\ref{fig:teaser}-B3e). Clicking on a bar takes users to the relevant conversation content (Fig.~\ref{fig:teaser}-C2a).
The filtering function (Fig.~\ref{fig:teaser}-B3c) allows users to zoom in on specific periods, magnifying the bars for more detailed analysis. 
\RR{The progress bar consolidates behaviors, minimizing visual clutter and highlighting key shifts in temporal behavior in the overview, while also allowing for detailed tracking of group dynamics (Fig.~\ref{fig:group18}-c).}

Three line charts track the behaviors and cognitive engagement of three students in the group (Fig.~\ref{fig:teaser}-B3b). Engagement is calculated at the midpoint and end of each question, and Savitzky–Golay filtering smooths the curves to highlight dynamic changes and trends across questions.
Below the charts, role types are mapped using equally sized rectangles at each timestamp. Users can zoom in on the timeline with the progress bar for detailed analysis.

\subsection{Detail View}
In the Timeline Panel, users can link to specific conversation content to review group discussions (Fig.~\ref{fig:teaser}-C1) and individual student conversations for each question (Fig.~\ref{fig:teaser}-C2). The playback feature allows users to revisit the original video of collaborative programming sessions, providing a more immersive classroom experience. This feature validates analysis results, offers detailed references, and supports the final step of our analysis workflow (R6).

\begin{figure*}
 	\centering
 	\includegraphics[width=1\linewidth]{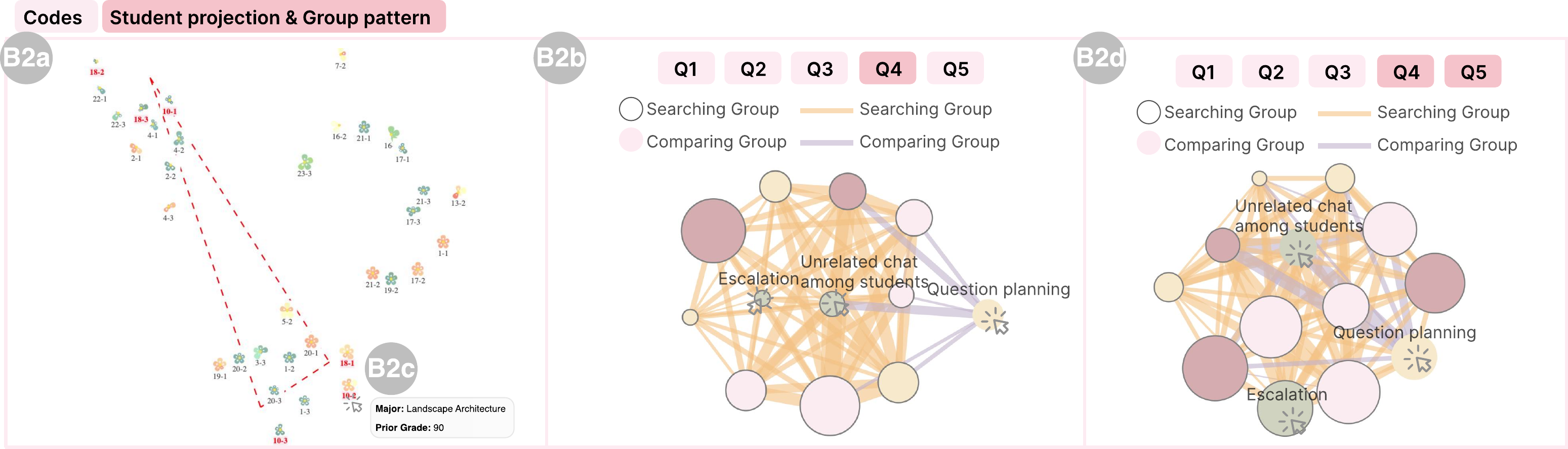}
 	\vspace{-1em}
 	\caption{The panel displayed in the Content View after selecting ``Student Projection \& Group Pattern'' is shown \RR{(A screenshot of the Group 10 \& 18 view)}. On the left (B2a), the projection of all students highlights the searched and compared groups, showing connections between students within the searched group. Hovering over a flower reveals the student's background information (B2c). On the right (B2b), group patterns for two groups across different questions are shown, with clickable question buttons to explore changes in dynamic learning behavior.}
        \Description{The panel displayed in the Content View after selecting ``Student Projection \& Group Pattern'' is shown \RR{(A screenshot of the Group 10 \& 18 view)}. On the left (B2a), the projection of all students highlights the searched and compared groups, showing connections between students within the searched group. Hovering over a flower reveals the student's background information (B2c). On the right (B2b), group patterns for two groups across different questions are shown, with clickable question buttons to explore changes in dynamic learning behavior.}
 	\label{fig:student} \end{figure*}

\section{Evaluation}
\label{evaluation}
\RR{To comprehensively evaluate the effectiveness of \textit{CPVis} in analyzing collaborative programming, we adopted a multi-faceted research approach. First, we conducted a quantitative study to validate the accuracy of collaboration performance annotations generated by LLMs. Next, we demonstrated two cases explored by computer science professors to showcase how \textit{CPVis} supported group evaluations and provided personalized feedback to students.
Finally, we recruited 22 participants for a user study to assess the practicality and user experience of \textit{CPVis}.}
All study designs were approved by our university's IRB.

\RR{\subsection{Quantitative Study}}
\RR{Using an LLM-driven approach for data annotation may raise trust concerns~\cite{liao2023ai}. To verify whether LLM performance impacts the validity of \textit{CPVis} results, we conducted a quantitative study to assess the accuracy and reliability of this method by sampling 20\% of the dataset. Specifically, we analyzed the following areas: evaluating code quality, annotating collaborative programming behaviors, identifying student roles, and recognizing teacher scaffolding.}

Inspired by CFlow~\cite{zhang2024cflow}, we recruited two experienced collaborative programming instructors (I1-I2) to annotate the data in four aspects. For instance, in the code quality evaluation, we asked the instructors to grade based on the same criteria used by ChatGPT-4o, \RR{with a score ranging from 1 to 5 for each dimension.} Detailed scoring criteria can be found in Sec.~\ref{code}.
First, I1 reviewed ChatGPT-4o's annotations and updated their original labels if necessary. I2 was then tasked with comparing I1's revised results (Version 1) and ChatGPT-4o's results (Version 2) without knowing the source of either label set.
\RR{I2 had to select one of the following four options: (1) I agree with Version 1, (2) I agree with Version 2, (3) Both versions seem valid, or (4) Neither version seems valid.}
We evaluated LLMs' performance in code quality by comparing it to human-labeled results.
The results showed that I1 and I2 reached 93.43\% agreement, while ChatGPT-4o 's annotations matched I1 and I2's annotations with 85.62\% and 89.32\% consistency, respectively. Therefore, we concluded that ChatGPT-4o's annotations were reliable, and we trusted its results.
Additionally, we employed the same method to assess three other dimensions.
Among these, ChatGPT-4o's accuracy was relatively lower in classifying collaborative programming behaviors (90.32\%) and code quality (93.43\%) but higher in identifying student roles (96.54\%) and teacher scaffolding (97.42\%). 
\RR{To mitigate the impact of annotation errors, we added prediction percentage and explanations to ChatGPT-4o's annotations of collaborative programming behaviors (Fig.~\ref{fig:teaser}-B3e), indicating the uncertainty of classification.} Similarly, we added explanations for code quality (Fig.~\ref{fig:teaser}-B1), providing more evidence for instructors during analysis.

\begin{figure*}
	\centering
	\includegraphics[width=1\linewidth]{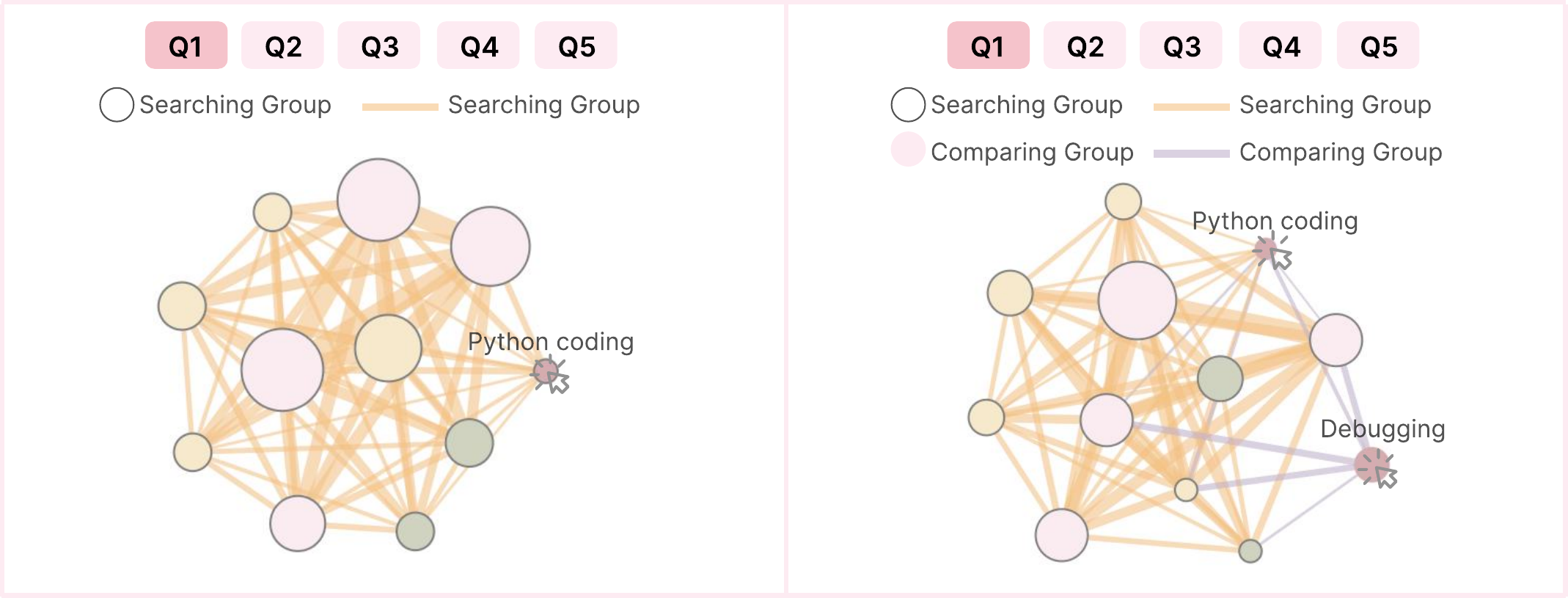}
	\caption{\RR{A screenshot of the ``Group Pattern'' view for Group 10 and Group 18,} with the left side showing Group 10's behavioral pattern in Q1 and the right side displaying the results after adding the comparison group 18. In comparison, it is evident that Group 10's ``Python coding'' behavior decreased, while Group 18's ``Debugging'' behavior increased, indicating more in-depth programming discussions in Group 18 during Q1.}
        \Description{A screenshot of the ``Group Pattern'' view for Group 10 and Group 18, with the left side showing Group 10's behavioral pattern in Q1 and the right side displaying the results after adding the comparison group 18. In comparison, it is evident that Group 10's ``Python coding'' behavior decreased, while Group 18's ``Debugging'' behavior increased, indicating more in-depth programming discussions in Group 18 during Q1.}
	\label{fig:network} \end{figure*}

\subsection{Case Study}
\label{case}
\RR{
We re-invited four computer science professors (P1-P4) from three universities who teach relevant courses and have extensive expertise in collaborative programming and data visualization to evaluate \textit{CPVis} and explore a real-world dataset using the system. 
We began by introducing the system's background, visual design, and workflow demonstration. Through two case studies based on the professors' interactions with \textit{CPVis}, we showcased the system's effectiveness in evaluating group performance and providing personalized feedback to students.
}
\subsubsection{Case 1: Assessing Similar Groups' Performance}
\begin{figure*}
    \centering
    \includegraphics[width=1\linewidth]{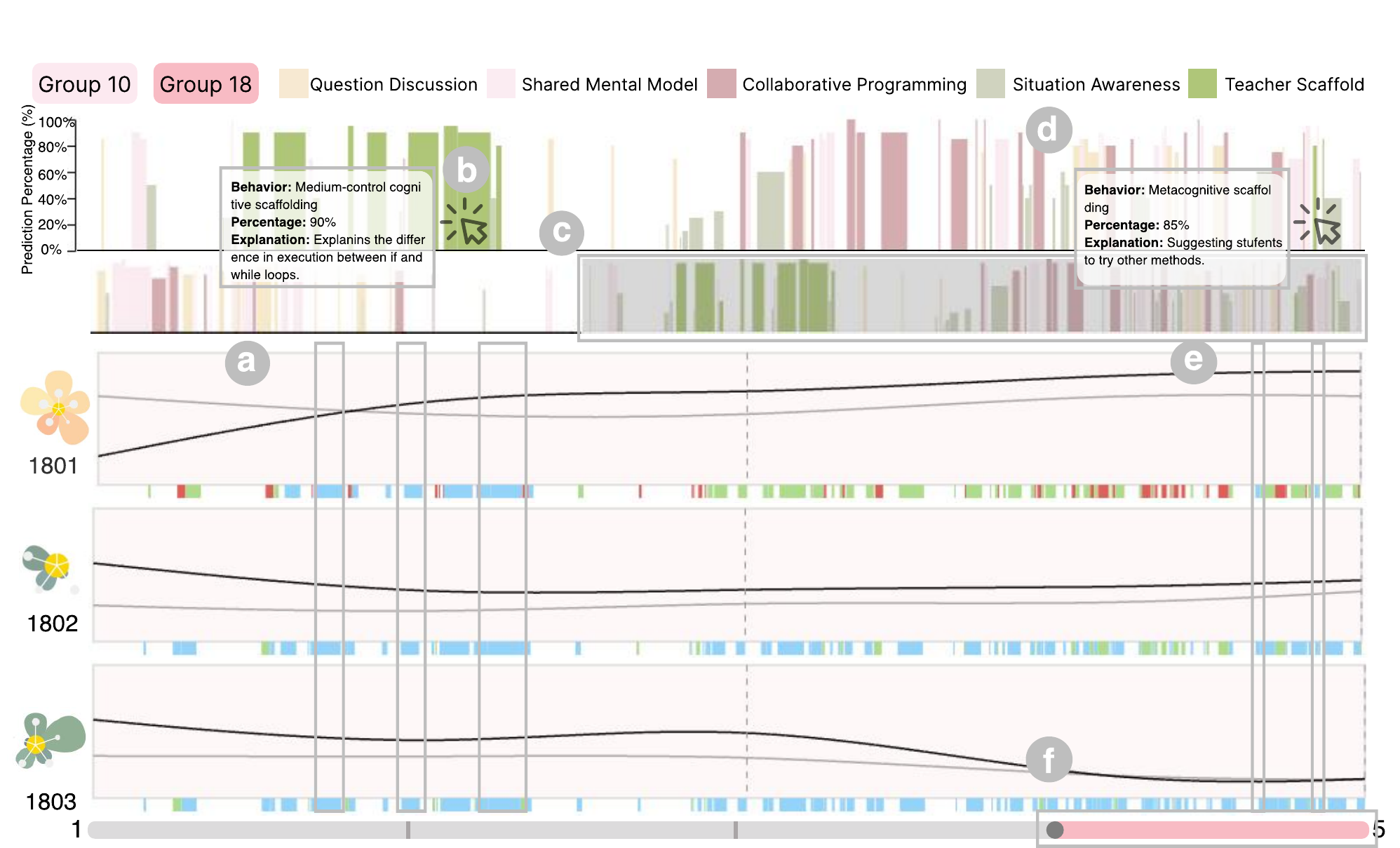}
    \caption{A screenshot of Group 18: (a) \& (e) show all three students acting as monitors while the instructor is speaking. (b) \& (d) indicate that the current bar represents the instructor's scaffolding provided, with a tooltip added. (c) \& (f) show that Q4-Q5 were selected on the progress bar.}
    \Description{A screenshot of Group 18: (a) \& (e) show all three students acting as monitors while the instructor is speaking. (b) \& (d) indicate that the current bar represents the instructor's scaffolding provided, with a tooltip added. (c) \& (f) show that Q4-Q5 were selected on the progress bar.}
    \label{fig:group18}
\end{figure*}

\RR{
The case summarized from P1 and P3 focused on the evaluation of group performance in collaborative programming and the feedback provided.}
They began by examining the filter view, where two closely positioned triangles with darker colors and larger outer arcs caught their attention (Fig.~\ref{fig:teaser}-A1b). 
Interested in comparing the two groups, they used the lasso tool to select them for further analysis.
The bouquet visualization revealed that Group 10 had one more butterfly than Group 18 (Fig.~\ref{fig:teaser}-A2), indicating stronger collaborative problem-solving skills.
Group 10's petals were more complete, and its leaves were greener, suggesting higher student engagement and a greater level of teacher scaffolding. 
\RR{Next, they searched for Group 10 (Fig.~\ref{fig:teaser}-A1a) and noticed that Group 18 was identified as the most similar group in the similarity panel (Fig.~\ref{fig:teaser}-A3). 
They selected Group 18 for comparison.}
In the code panel (Fig.~\ref{fig:teaser}-B1), they observed differences in the solutions for Q1 and Q5, while the rest of the questions were identical.
Coding analysis highlighted that Group 18 failed to output List A as required in Q1, resulting in a lower score.
For Q5, Group 18 used the \textit{`eval()'} function, while Group 10 correctly used \textit{`int()'} for type conversion, avoiding potential security risks and scoring higher (Fig.~\ref{fig:teaser}-B1).
While exploring the behavior pattern view (Fig.~\ref{fig:student}-B2b), they observed a yellow node in Group 18 gradually enlarging, representing ``Question Planning'' (Fig.~\ref{fig:student}-B2d).
Curious, they switched to the timeline panel and filtered for the discussion periods corresponding to Q4 and Q5 (Fig.~\ref{fig:group18}-c\&f).
Through the timeline, they discovered that during a specific period in Q4, all participants in Group 18 were assigned the ``Monitor'' role, indicated by blue bars (Fig.~\ref{fig:group18}-a).
By clicking on the bar, they found that the instructor had intervened, offering \textit{medium-control cognitive scaffolding} to guide students in adjusting their code and encouraging persistence (Fig.~\ref{fig:group18}-b).
\RR{For Q5, the instructor's involvement was shorter (Fig.~\ref{fig:group18}-e), providing \textit{metacognitive scaffolding} to encourage the group to try alternative approaches (Fig.~\ref{fig:group18}-d), such as using \textit{`eval()'}.}
Ultimately, Group 18 followed the instructor's advice and experimented with \textit{`eval()'}.
Through the timeline panel (Fig.~\ref{fig:teaser}-B3d), they found that the instructor provided \textit{high control cognitive scaffolding} (Fig.~\ref{fig:teaser}-B3e) for Q1 and Q4 of the group 10 and explained the detailed solutions (Fig.~\ref{fig:teaser}-C2a).
Finally, they validated these observations by reviewing classroom recordings.
They concluded that although Group 10 submitted higher-quality code, it benefited from greater instructional support. Therefore, Group 18 demonstrated better overall performance, considering its lower reliance on teacher scaffolding.

\begin{figure*}
	\centering
	\includegraphics[width=1\linewidth]{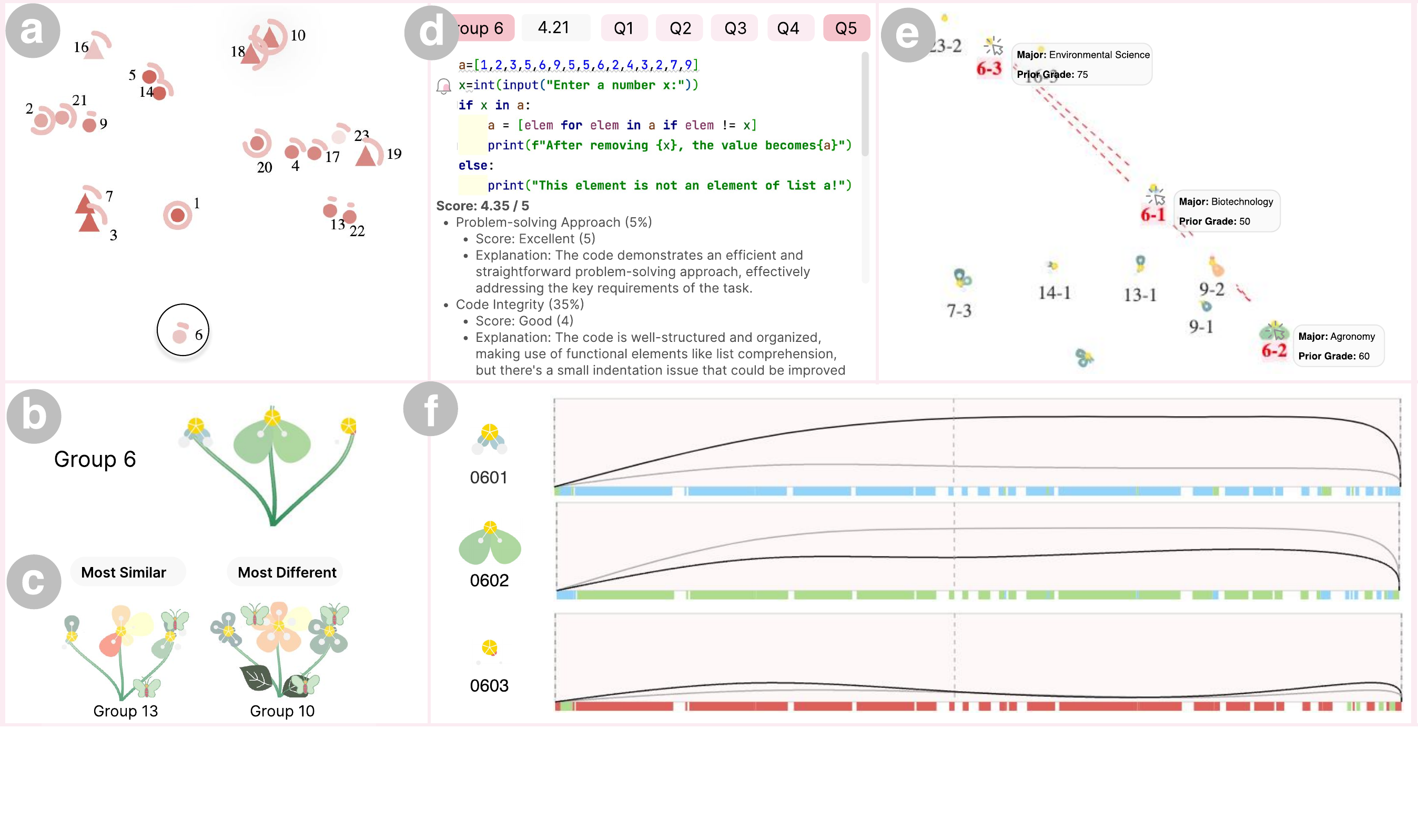}
	\caption{A screenshot of Group 6: (a) the Filter View, (b) the Group 6's bouquets, (c) the similarity panel, (d) the codes panel, (e) the student projection panel, and (f) the timeline panel.}
	\Description{A screenshot of Group 6: (a) the Filter View, (b) the Group 6's bouquets, (c) the similarity panel, (d) the codes panel, (e) the student projection panel, and (f) the timeline panel.}
	\label{fig:group6}
\end{figure*}

\subsubsection{Case 2: Providing Personalized Feedback for Students.}
\RR{The case summarized from the exploration process of P2 and P4 focused on providing personalized feedback, with our system helping identify both disengaged students and highly engaged ones who need further support.}
Using the lasso tool in the filter view (Fig.\ref{fig:group6}-a), the professors identified flowers missing petals and with smaller stamens (Fig.\ref{fig:group6}-b). Group 6's particularly small petals prompted further exploration. In the code panel, they noticed Group 6 consistently received scores above 4 (Fig.\ref{fig:group6}-d), indicating strong performance. 
However, in the timeline panel, they saw that the students' roles—Monitor (0601), Navigator (0602), and Driver (0603)—were independent with minimal collaboration (Fig.~\ref{fig:group6}-f). They also observed incomplete engagement data, with three missing questions. After reviewing the video, the professors confirmed the lack of collaboration and suggested encouraging more active engagement and collaboration among all members.

\RR{They noticed that Group 10 was the most different from Group 6 in the similarity panel (Fig.~\ref{fig:group6}-c). }
By selecting Group 10 in the filter panel, they observed that Group 10 had three butterflies (Fig.~\ref{fig:teaser}-A2), indicating a high level of collaborative problem-solving skills. 
Within the group, student 1002 stood out, with their flower being orange and having larger stamens, suggesting high levels of both behavioral and cognitive engagement. In the student projection panel, it was revealed that 1002 had a prior score of 90 (Fig.~\ref{fig:student}-B2a). 
On the timeline panel, 1002's curve was significantly higher than those of the other group members, confirming a high level of engagement. This student primarily assumed the role of Driver and frequently switched roles during discussions.
\RR{By exploring the timeline view, it was found that in Q5, 1002 repeatedly sought help from the instructor while other group members were discussing. Finally, the instructor provided high-control cognitive scaffolding (Fig.~\ref{fig:teaser}-B3d) to explain in detail how to solve the problem and pointed out a syntax error made by the group.}
Through their analysis, the professors discovered that although 1002 was actively engaged, they tended to rely excessively on the instructor during group discussions.  
The professors recommended that 1002 focus on developing independent thinking skills by first consulting resources or discussing with group members when facing challenges rather than immediately seeking instructor assistance.

\subsection{User Study}
\RR{User Study focused on three core objectives: G1, to assess \textit{CPVis}'s performance in collaborative group evaluation tasks; G2, to validate the effectiveness of \textit{CPVis} in delivering personalized feedback to students; and G3, to evaluate the overall user experience of \textit{CPVis}. }
\subsubsection{Participants}

We re-recruited 10 instructors and 12 teaching assistants (TAs) from the Computer Science department to participate in this user study (P1-P22, age: $ 28.95 \pm 5.21 $). Among the instructors, 2 were professors, and 8 were assistant professors, with 6 male and 4 female participants. On average, the instructors had 4.6 years of experience teaching programming courses, while the TAs had an average of 1.5 years of experience assisting with programming courses. All participants had experience teaching or co-teaching programming courses and were familiar with collaborative programming education. 
Given the study's focus on collaborative programming environments, we only recruited participants with a computer science background to ensure they could provide authentic and interpretable insights for our research. All sessions were conducted online via Zoom, and each participant received a \$20 reward at the end of the study.

\subsubsection{Study Design and Procedure}

We conducted a within-subjects user study comparing the full version of \textit{CPVis} with two baseline systems. 

\RR{
\textbf{Baseline System 1:} In real-world classroom settings, instructors typically evaluate students and groups based on raw data. 
To simulate traditional post-class evaluations, baseline system 1 provided only raw data, including students' background information (such as major and past grades), group-submitted code, and group discussion videos. The design of Baseline System 1 aimed to assess whether visualizing collaborative programming data could improve the efficiency and accuracy of instructors' evaluations of student and group performance.

\textbf{Baseline System 2:} Since no systems were available for multi-level assessment of collaborative programming in real classrooms, we selected the full \textit{CPVis} system for an ablation study. Baseline System 2 retained the same learning analytics features and visual design as \textit{CPVis} but excluded the comparison feature. The design of Baseline System 2 aimed to assess whether adding more panels introduced additional complexity and whether the comparison feature in \textit{CPVis} improved the efficiency of instructors' assessments.
}

We randomly selected 10 groups from 19 as experimental datasets for each system, ensuring that the data for each system was entirely distinct.
Participants were also randomly assigned to experiment with all three systems, ensuring fairness and randomness in the study.
The actual experiment consisted of three phases:

\textbf{\textit{Introduction:}} First, we introduced participants to the primary objectives of the system and the research background through a video. After obtaining their informed consent and collecting their personal information, we provided a detailed explanation of the key concepts and metaphors involved in the experiment to ensure participants could understand and effectively utilize these metaphors in the following phases. By administering a basic metaphor comprehension questionnaire (see supplementary materials), we confirmed that participants had fully grasped the necessary metaphors. The entire introduction phase lasted approximately 20 minutes.

\textbf{\textit{Exploration:}} In this phase, participants were asked to explore three systems, perform assigned tasks, and explore freely or based on their own needs. Each participant was required to use three systems to complete two tasks. The first task involved rating the collaborative groups in the dataset (2 as excellent, 2 as good, 2 as fair, 2 as pass, and 2 as fail). The second task was to identify students who needed personalized feedback. To prevent the task completion time from being influenced by the order, we shuffled the group IDs in each system to balance the difficulty across the two systems, ensuring fairness in the experimental results.

\begin{figure*}
	\centering
	\includegraphics[width=1\linewidth]{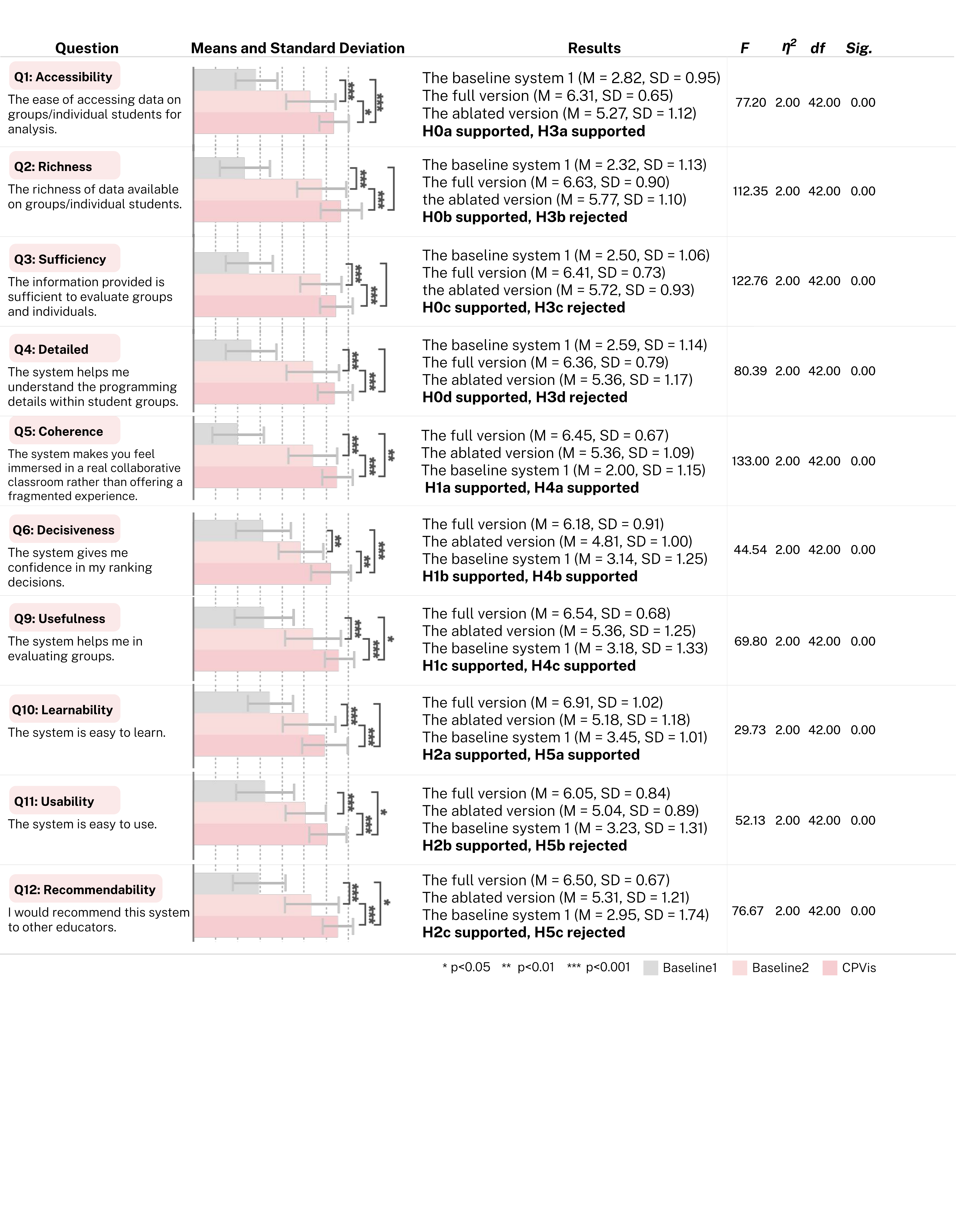}
	\caption{Results of the RM-ANOVA for the Baseline 1, Baseline 2, and full version \textit{CPVis} questionnaires, based on a 7-point Likert scale (1 means disagree, 7 means agree), including the Mean (M) and Standard Deviation (SD).}
        \Description{Results of the RM-ANOVA for the Baseline 1, Baseline 2, and full version \textit{CPVis} questionnaires, based on a 7-point Likert scale (1 means disagree, 7 means agree), including the Mean (M) and Standard Deviation (SD).}
	\label{fig:result}
\end{figure*}

\textbf{\textit{Feedback:}} Based on the participants' experience with the system, they were asked to complete a user experience questionnaire, which included 7-point Likert scale questions derived from existing literature~\cite{xia2019peerlens,shi2018meetingvis}, \RR{as shown in Fig.~\ref{fig:result}.} Additionally, we conducted follow-up interviews to gather more feedback and insights about our system and its potential real-world applications. These interviews lasted approximately 15 minutes.

\textbf{\textit{Hypotheses:}}
We propose the following hypotheses based on Peerlens~\cite{xia2019peerlens}:

\textit{H0: }Both the full and ablated versions of \textit{CPVis} outperform Baseline System 1 in terms of information access and analysis. Specifically, \textit{CPVis} shows advantages in accessibility (H0a), data richness (H0b), sufficiency (H0c), and detailed (H0d) compared to Baseline System 1.

\textit{H1: }Both the full and ablated versions of \textit{CPVis} outperform Baseline System 1 in terms of user experience and decision support. Specifically, \textit{CPVis} excels in coherence (H1a), decisiveness (H1b), and usefulness (H1c) over Baseline System 1.

\textit{H2: }Both the full and ablated versions of \textit{CPVis} surpass Baseline System 1 in terms of ease of use and recommendability. Specifically, \textit{CPVis} excels in learnability (H2a), usability (H2b), and recommendability (H2c) compared to Baseline System 1.

\textit{H3: }The full version of \textit{CPVis} provides more information than the ablated version. Specifically, the full version outperforms the ablated version in accessibility (H3a), data richness (H3b), information sufficiency (H3c), and detailed (H3d).

\textit{H4: }The full version of \textit{CPVis} outperforms the ablated version in user experience and decision support. Specifically, the full version excels in coherence (H4a), decisiveness (H4b), and usefulness (H4c).

\textit{H5: }The ablated version of \textit{CPVis} is superior to the full version. Specifically, the ablated version is perceived as more learnable (H5a), more usable (H5b), and more recommendable (H5c) compared to the full version.
\subsection{Result and Analysis}
\RR{
We reported the evaluation results based on two completed tasks and the overall system experience. A repeated measures analysis of variance (RM-ANOVA) was conducted for each questionnaire item, followed by Bonferroni post-hoc tests for measures with significant differences. The results showed favorable performance and positive feedback on G1, G2, and G3. The ratings are shown in Fig. ~\ref{fig:result}.
}

\textbf{\textit{G1, to assess \textit{CPVis}'s performance in collaborative group evaluation tasks.}}
Participants completed the task of evaluating collaborative groups but varying evaluating standards made it challenging to quantify the accuracy of their evaluations. To understand their criteria, we asked, ``What criteria did you use to rate these groups?''
In \textit{Baseline 1}, participants evaluated factors such as code errors, meeting the question requirements, code quality, time complexity, readability, and the use of provided methods. 
In \textit{Baseline 2}, more learning analytics and visualization allowed participants to expand their evaluation criteria. 
They focused on ChatGPT's code assessment, student engagement, quality of collaborative problem-solving, student roles, and teacher scaffolding, as represented in the flower-based visualization. 
\textit{P5 mentioned that, due to similar group answers, assessing accuracy from code alone was challenging. However, the flower-based visualization helped reveal both group and individual performance.}
In \textit{CPVis}, the introduction of comparison panels further expanded the evaluating criteria. \textit{P13 noted that the behavior pattern view in the comparison panel played a crucial role in their evaluations, as it allowed them to analyze behavior patterns over time for two groups within the same interface. This was particularly useful for comparing highly similar groups.}

\RR{
Regarding time spent on the task, we found that although \textit{Baseline 1} provided a large amount of data, it was relatively raw. Participants typically lacked the patience to examine all the details and relied more on code scoring, spending an average of 20 minutes.}
In contrast, participants using \textit{Baseline 2} and \textit{CPVis} systems spent more time, averaging 35 minutes and 30 minutes, respectively. We believe this was due to the richer learning analytics provided in these versions, which engaged participants more deeply in exploring collaborative programming details.

\textbf{\textit{G2, to assess the effectiveness of \textit{CPVis} in providing personalized feedback to students.}}
We gained valuable insights by asking participants, ``Which two students do you think need feedback the most, and why?''
In \textit{Baseline 1}, since the code was submitted as a group, participants found it challenging to assess individual performance, making it difficult to provide personalized feedback.
In \textit{Baseline 2}, participants quickly identified students needing feedback, focusing on outliers with low engagement or imbalanced roles. Feedback typically encouraged more active engagement in discussions and collaboration.
In \textit{CPVis}, participants highlighted students with ``contradictory'' traits. \textit{P8 pointed out a student with a poor academic record who displayed high behavioral engagement in the current group programming project, clearly trying to keep up with the group. Conversely, P9 identified a student with a strong academic record who showed low engagement and remained mostly silent during discussions.}
\RR{We believe that \textit{CPVis}'s visualization capabilities not only help participants quickly identify students with unusual performance but also allow them to recognize students who may require additional attention or support.}

\textbf{\textit{G3, to assess the overall user experience of \textit{CPVis}. }}
Figure~\ref{fig:result} presented the questionnaire results. \RR{Overall, the ablated and full versions of \textit{CPVis} scored significantly higher than Baseline1 across all evaluation metrics, with statistically significant differences. 
All metric scores of the full version of \textit{CPVis} were higher than those of the ablated version (Baseline 2).
Compared to the ablated version, the full version of \textit{CPVis} introduced comparison features, resulting in more panels. However, the results showed that, except for \textit{Richness}, \textit{Sufficiency}, \textit{Detailed}, and \textit{Learnability}, all other metrics exhibited statistically significant differences. 
Thus, hypotheses \textit{H3b, H3c, H3d, H5b, and H5c} were not supported, while \textit{H3a, H4a, H4b, H4c, and H5a} were supported.
It indicated that the full version of \textit{CPVis} added valuable features without introducing unnecessary complexity. 
Although no significant differences were observed in the \textit{Learnability} metric, both the full and ablated versions were easy to learn and effectively enhanced the instructor's assessment efficiency.}
\textit{P10 stated, ``The group behavior pattern analysis in the full version helped me a lot. Comparing the two groups made it easier to see which group had more in-depth collaboration. In contrast, directly comparing two node connection diagrams would have been much harder.''}
Additionally, participants consistently recognized the value of learning analytics during follow-up interviews. \textit{CPVis} provided multi-level learning analysis from groups to individual students. While the ablated version reduced some comparison features, it did not remove any analytical dimensions, which explained the lack of significant differences in \textit{Richness}, \textit{Sufficiency}, and \textit{Detailed}. 
\RR{On average, participants still gave higher ratings to \textit{CPVis}.}
Participants provided separate ratings for these designs in Q7 and Q8, with scores ranging from 5 to 7. For the \textit{Visual Encoding} question ($M = 6.63, SD = 0.58$) \raisebox{-0.2\height}{\includegraphics[width=0.5cm]{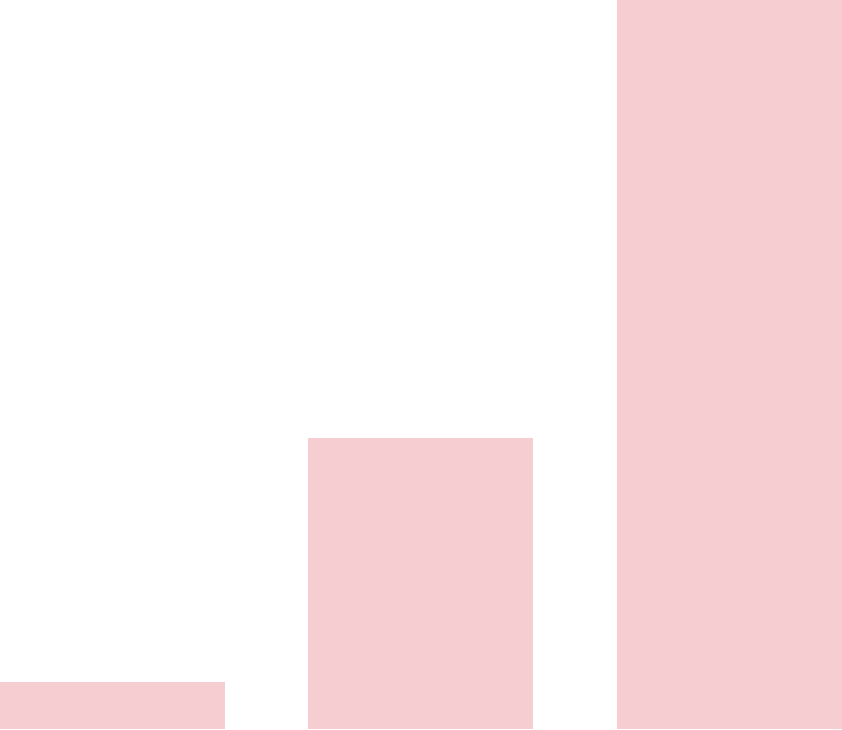}}, most participants found the flower metaphor intuitive and thought the visual design was highly innovative. 
P14 remarked that she felt a sense of familiarity with the flower design, stating that using flowers to represent students was easy to understand and accept, with a reasonable encoding method.
For the \textit{Visual Design} ($M = 6.32, SD = 0.78$) \raisebox{-0.2\height}{\includegraphics[width=0.5cm]{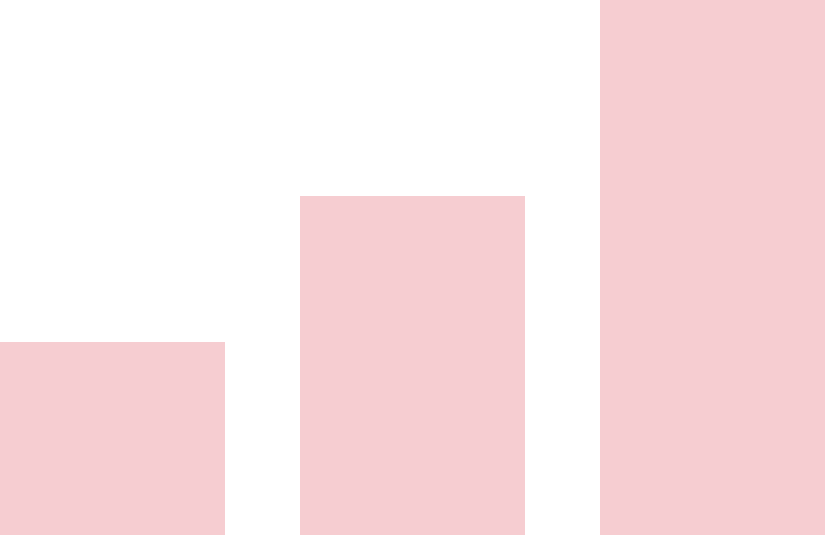}}, participants noted that by observing the design of flowers or bouquets, they could quickly and intuitively understand the overall situation of groups and students, which facilitated further exploration of details.

\section{Discussion}
\RR{
Our study utilizes an intuitive flower-based visual design and evidence-based collaborative programming process analysis to provide instructors with a clear perspective for evaluating group and individual performance in collaborative programming. In this section, we discuss the lessons learned, the factors contributing to the research outcomes, and how these findings relate to existing works.

\subsection{Flower-Based Visual Design for Intuitive and Useful by Participants}
In large-scale learning analytics, intuitive visualization and interactive features prove to be valuable in assisting instructors with evaluations while reducing their workload~\cite{martinez2020data,fernandez2024data}.
Our study shows that the flower-based visual design effectively helps instructors summarize the performance of students and groups in collaborative programming.
Participants using \textit{CPVis} typically report starting by observing the flower visualization to gain an overview of the group's overall performance and the engagement levels of individual members during collaboration. Our design enables them to make quick assessment judgments and uncover valuable educational insights. 
For instance, students playing the Driver role often exhibit higher engagement levels.
\RA{Previous works use dynamic natural metaphors~\cite{tausch2014groupgarden,tausch2016comparison}, such as blooming flowers, falling leaves, and weather changes, to represent the quality and state of group discussions. However, these metaphors primarily convey overall trends or atmospheres rather than offering a precise and structured representation of multidimensional data, making it difficult for users to extract specific and accurate information efficiently. Moreover, the strong symbolic and emotional nature of their metaphors often leads to subjective interpretations.}
The effectiveness of our design lies in its ability to translate multiple dimensions of process-based learning analytics into visual elements such as colored petals and flower stamens, enabling instructors to quickly interpret multidimensional data and assess both group and individual performance during collaboration.
Furthermore, the flower-based visualization supports hierarchical analysis at both the group and individual levels, allowing instructors to efficiently analyze and compare the performance of multiple groups and students on a large scale.

\subsection{\textit{CPVis} Enhanced Instructors' Confidence in Evaluating Groups and Students}
The study demonstrates that \textit{CPVis} enhances participants' confidence in evaluation outcomes and improves the accuracy of their assessments. In Baseline System 1, participants report that accessing data requires significant time, and evaluating a specific group's performance often necessitates finding similar groups for a relatively fair comparison. 
Such a process demands additional time, causing participants to lose patience and avoid thoroughly examining all the details.
In baseline system 2, participants have to manually browse and process large amounts of student behavior and interaction data, which significantly increases cognitive load and reduces efficiency as they rely on memory to evaluate the performance of different groups.
In comparison, \textit{CPVis} offers significant convenience to participants \RA{by visualizing multidimensional learning analytics data}, allowing them to effortlessly access key information required for evaluations and compare similar groups. By providing both an overall view of multiple groups and detailed comparisons into individual groups, \textit{CPVis} substantially boosts participants' confidence in their evaluation outcomes, as demonstrated in the ratings. 
\RA{Clear and intuitive visual analytics systems contribute to improved confidence and efficiency among participants. For instance, Groupnamics helps participants identify groups requiring intervention by visualizing each group's recent vocal activities and discussion statuses in a one-page view, thereby boosting their confidence in decision-making~\cite{sato2023groupnamics}.}
While it is ideal for \textit{CPVis} to support comparisons across an unlimited number of groups, practical limitations related to cognitive load and visual design make this challenging. Future efforts focus on optimizing the evaluation process through visual design, striking a balance between cognitive load and evaluation efficiency, thereby providing effective support for teaching.

\subsection{Theory-driven and LLM-powered Automation Evaluation for Quantifying Collaborative Learning}
Our study utilizes data collection, analysis, and visualization techniques to extract key insights from students' collaborative behaviors and outcomes, providing a deeper understanding of the learning process in collaborative programming. We focus on quantifying complex collaborative learning processes by leveraging LLMs and theoretical frameworks, introducing innovative methods to evaluate collaboration efficiency. 
While collaborative problem-solving is clearly defined in prior research~\cite{rosen2020towards}, achieving a quantitative balance between task performance and team effectiveness remains a significant challenge. To address this, we employ the coefficient of variation as a balancing metric and validate its efficacy using real-world datasets.
By integrating LLMs, \textit{CPVis} automates the annotation of collaborative programming performance, significantly reducing the workload associated with manually labeling large-scale classroom data and offering a novel perspective for automated learning analytics. 
Combining theory-driven metrics and LLM-powered automation provides instructors with robust, multidimensional evidence, enabling them to process and compare extensive student data systematically. 
This empowers instructors to effectively evaluate group and individual behaviors in collaborative programming, identify collaboration patterns, and support evidence-based decision-making. Previous research demonstrates that data-driven analysis helps educational decision-makers~\cite{hou2024codetailor}, such as instructors, uncover hidden learning patterns and deliver personalized guidance. Building on this foundation, \textit{CPVis} further enhances the potential for personalized feedback, enabling instructors to provide precise, data-driven guidance to students.

}

\section{Limitations and Future Work}

\RR{
In this section, we discuss the limitations of the current study and potential future work.
\subsection{Limitation}
Our study has three main limitations.
First, our current analysis is limited to data from a single real-world classroom's collaborative programming discussions, restricting the generalizability of our findings to other contexts. Similarly, our evaluation of \textit{CPVis} relies on a sampled dataset, limiting the study's scope. We hypothesize that participants working with smaller datasets and visualized learning analytics experience reduced cognitive load and find it easier to identify collaboration patterns due to fewer visual elements to process. However, in large-scale collaborative programming classrooms, instructors face the challenge of evaluating more groups and students, which may increase memory load and visual complexity.
Second, the data collected in our study are obtained from real classroom environments, maintaining ecological validity by capturing natural behaviors such as group silence or requests for instructor assistance. However, due to the limitations of non-intrusive equipment, our data lack details such as facial expressions and non-verbal cues. While participants report the comprehensiveness and richness of the learning analytics in the experiment, the absence of these data poses challenges for deeper analysis of emotional expressions and social engagement during collaborative programming. This limitation hinders the provision of a more holistic learning analysis for evaluation purposes.
Additionally, the recorded data are independent and exclude audio information, making it difficult to align screen interactions with dialogue streams. This limitation constrains the exploration of the relationship between collaborative behavior patterns and collaborative problem-solving processes.
Finally, in large-scale collaborative programming classrooms, generating analytics using LLMs requires significant computational time and cost. While feasible for institutions with robust computational resources, this remains a limitation for deploying such tools in real teaching scenarios. Furthermore, in real classrooms, noise from multiple group discussions introduces significant data noise, complicating the automation of learning analytics generation and limiting the accuracy of evaluations for groups and individual students.

\subsection{Future Work}
Without well-structured visualizations, simply presenting multiple data streams poses significant challenges for instructors attempting to interpret these large-scale datasets~\cite{fernandez2024data}.
In this study, we explore the integration and analysis of multimodal data. However, \textit{CPVis} has the potential to further enhance the visualization and perception of multimodal data, enabling instructors to evaluate group and student performance with greater accuracy and reduced cognitive load~\cite{martinez2020data}. 
Our target audience consists of instructors teaching large introductory collaborative programming courses, who require more efficient and intuitive visualizations to understand student performance during collaboration.  
While our use of static 2D visualizations, such as high-dimensional flower glyphs, has been highly regarded by participants for boosting confidence and helping instructors quickly identify key features, we believe there is room for improvement in organizing visualization formats to enhance information transmission efficiency and the users cognitive experience.
For instance, incorporating narrative visualizations further streamlines the process by allowing instructors to generate composite evaluations based on their weighting of different collaboration performance dimensions~\cite{gratzl2013lineup}. 
Narrative visualizations enable instructors to delve into data details, organize learning analytics results along logical paths such as timelines, causality, or categories, and highlight key information~\cite{chen2019designing}. 
This approach mitigates visual overload caused by excessive data, significantly reduces the time and cognitive effort required for evaluation, and ultimately supports instructors in making better decisions and assessments.

\textit{CPVis} requires instructors to spend additional time after class to evaluate collaborative performance. In our study, most participants indicate during follow-up interviews that the extra time spent on evaluating students' collaborative performance is highly valuable for producing comprehensive assessments. They note that providing immediate evaluations during the collaboration process is unrealistic, as final assessments typically need a holistic consideration of task completion and group dynamics after class. However, there is a significant demand for real-time analysis tools to deliver timely, personalized feedback to students and offer appropriate instructional scaffolding during the collaborative process~\cite{tang2024sphere}.
Instructors frequently find themselves overwhelmed by the immediate needs of some students~\cite{yang2023pair}, unintentionally neglecting others. To address this issue, future work could explore the integration of LLMs to enable real-time monitoring and analysis of students' behavioral data—such as code submissions, error logs, and engagement levels. LLMs could automatically detect learning bottlenecks or collaboration issues, providing instant feedback on common problems to students. This would effectively reduce instructors' workload, allowing them to focus on complex or critical issues, and simplify classroom management tasks.
For instance, LLMs could summarize patterns in students' code submissions and generate a ``hotspot report'' identifying recurring issues across the class. They could also provide real-time collaborative performance analytics for different groups, enabling instructors to quickly gain a comprehensive understanding of overall class dynamics. Additionally, LLMs could assist in role allocation within groups, suggest strategies to improve team interactions, and identify potential conflicts or disengagement within collaborative teams.
LLM-powered tools automate evaluations and enable personalized feedback, bridging post-class assessments with in-class scaffolding to enhance teaching and learning in collaborative programming.}
\section{Conclusion}
Collaborative programming has been a common teaching strategy for instructors in large-scale programming courses.
We collected multimodal collaborative programming data from real-world settings and, after consulting with domain experts, developed a performance framework for evaluating collaborative programming groups and individual students. We visualized these performance metrics by introducing a novel flower metaphor-based design and built an interactive visualization system to dynamically analyze group collaboration behavior and track students' evolving engagement over time.
\RR{Finally, we conducted a quantitative study to evaluate the accuracy of annotation data labeling by LLMs,} with two cases demonstrating how our system helped instructors evaluate group performance and provide personalized feedback to students. Additionally, we organized a within-subjects experiment (N=22) comparing \textit{CPVis} with two baseline systems. The results indicated that participants gained more insights using our system and felt significantly more confident in evaluating group collaboration.

\begin{acks}
Meng Xia is the corresponding author.
The work was supported by the National Natural Science Foundation of China, (62422607, 62372411, 62036009) and the Zhejiang Provincial Natural Science Foundation of China.

\end{acks}


\appendix
\RR{
\newpage
\section{DETAILS OF FORMATIVE STUDY}
\subsection{Background Information of Participants}
The table~\ref{tab:info} presents information about the participants in our formative study.
}

\begin{figure}[htbp]
	\centering
	\includegraphics[width=0.8\linewidth]{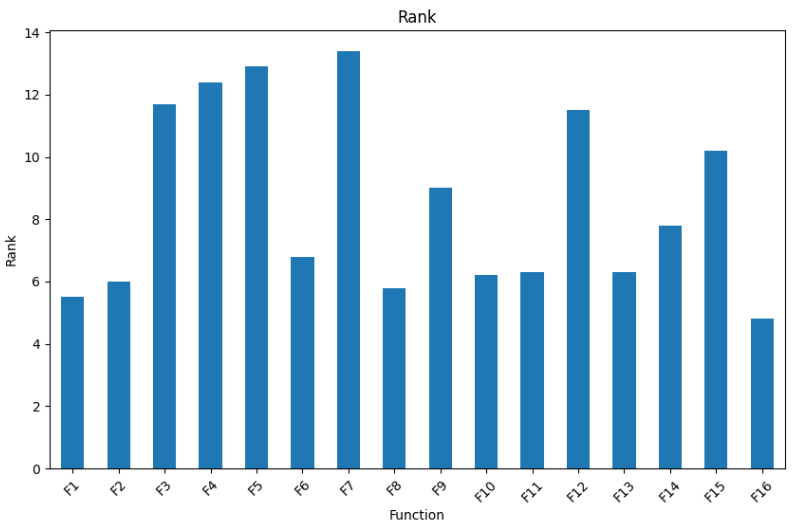}
	\caption{Ranking Results of F1-F16.}
        \Description{Ranking Results of F1-F16.}
	\label{fig:rank}
\end{figure}

\RR{
\subsection{Findings of the Semi-structured Interviews}

\begin{table*}[hb]
\renewcommand{\arraystretch}{1.5}
\centering
\begin{tabular}{m{1.5cm} m{1cm} m{1cm} m{1.5cm} m{3.5cm} m{2.5cm} m{3.5cm}}
\hline
Participants & Gender & Age & Teaching experience & Students taught & Whether the class is large-scale & Whether the students know each other before \\ \hline
E1 & Female & 45 & 15 & University Non-Computer Science Majors & Yes & No \\
E2 & Female & 30 & 5 & University Non-Computer Science Majors & Yes & No \\
TA1 & Male & 24 & 1 & University Non-Computer Science Majors & Yes & No \\
TA2 & Female & 24 & 1 & University Non-Computer Science Majors & Yes & No \\
T1 & Male & 31 & 3 & University Non-Computer Science Majors & Yes & No \\
T2 & Male & 29 & 5 & University Computer Science Majors & Yes & Yes \\
T3 & Male & 36 & 8 & University Computer Science Majors & No & Yes \\
T4 & Female & 26 & 3 & Vocational High School Non-Computer Majors & Yes & Yes \\
T5 & Female & 26 & 3 & Elementary School Students & No & Yes \\
T6 & Male & 35 & 7 & University Non-Computer Science Majors & Yes & No \\ \hline
\end{tabular}
\caption{The detailed background statistical information of the participants in the Formative Study.}
\label{tab:info}
\end{table*}

\textbf{\textit{Motivations:}}
Participants unanimously agreed that collaborative programming is an important teaching strategy for programming education. However, different teachers had varying focuses and practices regarding collaborative programming. 
Teachers specializing in computer science tended to have students complete tasks independently in foundational programming courses (e.g., Python, Java) while implementing collaborative programming for more complex projects. 
For example, \textit{T3 stated, ``I consider collaborative learning only when the programming project requires different students' creative input, innovation, exploration, and discovery of new content, or when the workload is too large for one person to complete.''}
In contrast, teachers of non-computer science courses held different views. 
E2 pointed out that for beginners who are new to programming languages, even a fundamental programming problem can be daunting.
Collaborative programming can significantly alleviate this issue and improve students' efficiency. 
Group members can work together to complete a programming task, which motivates students, allows them to learn from each other, and reduces the teaching burden (T4).
Additionally, in collaborative programming classes, teachers focus on students' cognitive skills (e.g., knowledge acquisition) and non-cognitive skills (e.g., computational thinking, problem-solving abilities). Most participants believed that these two aspects do not conflict and are part of the overall teaching goals. 
\textit{T5 stated that her course design primarily focuses on developing non-cognitive skills, particularly enhancing students' abilities during the collaboration process. }
T4 believed that cognitive skills can be supplemented with extra time or individual tutoring, but problem-solving and collaboration skills are challenging to develop outside of class. 
Furthermore, E1, T4, and T3 mentioned that quantifying the improvement of students' non-cognitive skills is very challenging.
On one hand, teachers and assistants are busy answering students' questions, making it challenging to observe students' performance. 
On the other hand, non-cognitive skills are not as quickly assessed through tests as knowledge-based content.

Despite the unanimous recognition of the importance of collaborative programming, participants shared several challenges they faced in practice, summarized as follows:

\textit{\textbf{Group Formation:}}
For group size, most participants preferred group sizes of 2-3 students, noting that larger groups (up to 5 members) could decrease the quality of collaboration and make it difficult for students to find their roles. T4 favored pair programming but often had to form groups of three due to large class sizes. T1 and T3 typically recommended groups of 4-5 students, given the complexity and workload of their programming projects, which smaller groups might struggle to complete.
For the group formation method, participants had various approaches to forming groups. T2, T3, T4, and T5, whose students come from fixed classes, generally allowed students to create their own groups. In contrast, T6, T1, E1, and E2 taught general education courses open to all students across the university. These students came from different majors and did not know each other, making self-selection difficult. E2 conducted preliminary assessments of students and formed balanced groups based on their prior knowledge. However, determining how to create groups remains a significant challenge, as different criteria can lead to varying outcomes. Arbitrarily formed groups might result in imbalanced teams, where students' abilities are insufficient to complete the tasks.

\textit{\textbf{Evaluating the Learning Objectives:}}
Evaluating whether the teacher has achieved the learning objectives is an essential criterion for assessing the quality of collaborative programming courses. All participants stated that they generally meet the learning objectives and sometimes even exceed expectations. T4 mentioned that at the start of the collaborative course, she assigned a group leader to each group to take on a leadership role. While she initially spent a significant amount of time guiding the group leaders and facilitating student collaboration, by the second half of the semester, the group leaders were typically able to function as ``little teachers,'' helping group members complete tasks. This allowed the teacher to free up time and energy for other responsibilities.

\subsection{Results of Rating and Ranking in Formative Study}
Table~\ref{table:feature} presents the metrics participants rated and ranked for the collaborative programming visual analytics system, as discussed in Section 3.2.}
\begin{table}[htbp]
\renewcommand{\arraystretch}{1.5}
\begin{tabular}{p{8cm}}
\hline
\textbf{Features}  \\ \hline
F1: Change in student engagement  \\
F2: Change in student behavior \\
F3: Analysis of student emotional state \\
F4: Student background statistics \\
F5: Analysis of student programming habits \\
F6: Change in role distribution \\
F7: Role assignment suggestions within the group \\
F8: Group performance overview \\
F9: Comparison of group performances \\
F10: Analysis of similarity in group performance  \\
F11: Generation of personalized suggestions \\
F12: View of raw data \\
F13: Programming task evaluation \\
F14: Visualization of problem-solving path \\
F15: Real-time feedback feature \\
F16: Students performance overview \\ \hline
\end{tabular}
\caption{Features of Collaborative Programming Visualization Systems Extracted from Semi-structured Interviews.}
\label{table:feature}
\end{table}
\RR{The raw rating data is displayed in Table~\ref{table:rating}.
The results show that participants rated most of the features mentioned highly, except for F3, F4, F5, F7, and F15.}
\begin{table*}[htbp]
\renewcommand{\arraystretch}{1.5}
\centering
\small
\begin{tabular}{l c c c c c c c c c c c c}
\hline
\textbf{Feature} & \textbf{T6} & \textbf{T5} & \textbf{E1} & \textbf{E2} & \textbf{TA1} & \textbf{TA2} & \textbf{T1} & \textbf{T2} & \textbf{T3} & \textbf{T4} & \textbf{Mean} & \textbf{SD} \\ \hline
F1 & 5 & 6 & 5 & 6 & 7 & 6 & 7 & 5 & 6 & 7 & 6 & 0.82 \\
F2 & 7 & 6 & 5 & 5 & 4 & 7 & 6 & 6 & 6 & 6 & 5.8 & 0.92 \\
F3 & 3 & 4 & 3 & 4 & 4 & 5 & 6 & 4 & 3 & 4 & 4 & 0.94 \\
F4 & 2 & 4 & 2 & 2 & 2 & 3 & 3 & 4 & 6 & 5 & 3.3 & 1.42 \\
F5 & 4 & 3 & 6 & 5 & 4 & 3 & 4 & 4 & 3 & 5 & 4.1 & 0.99 \\
F6 & 4 & 6 & 7 & 3 & 7 & 5 & 4 & 6 & 6 & 7 & 5.5 & 1.43 \\
F7 & 2 & 4 & 3 & 4 & 2 & 4 & 4 & 3 & 4 & 3 & 3.3 & 0.82 \\
F8 & 7 & 6 & 7 & 7 & 7 & 7 & 7 & 7 & 7 & 7 & 6.9 & 0.32 \\
F9 & 6 & 6 & 5 & 7 & 5 & 4 & 7 & 6 & 5 & 7 & 5.8 & 1.03 \\
F10 & 7 & 6 & 5 & 6 & 5 & 5 & 5 & 5 & 6 & 7 & 5.7 & 0.82 \\
F11 & 7 & 6 & 5 & 5 & 5 & 5 & 6 & 7 & 7 & 7 & 6 & 0.94 \\
F12 & 3 & 5 & 5 & 5 & 5 & 5 & 6 & 7 & 4 & 3 & 4.8 & 1.23 \\
F13 & 7 & 6 & 7 & 7 & 7 & 7 & 7 & 7 & 6 & 6 & 6.7 & 0.48 \\
F14 & 6 & 7 & 6 & 6 & 6 & 5 & 4 & 5 & 6 & 6 & 5.7 & 0.82 \\
F15 & 2 & 3 & 4 & 3 & 2 & 5 & 6 & 7 & 7 & 4 & 4.3 & 1.89 \\
F16 & 7 & 7 & 6 & 6 & 5 & 7 & 7 & 5 & 6 & 6 & 6.2 & 0.79 \\ \hline
\end{tabular}
\caption{Mean and SD Results of the Ratings for F1-F16 by Ten Participants.}
\label{table:rating}
\end{table*}
\RR{The raw ranking data is displayed in Table~\ref{table:rank}.
Each feature's average ranking is listed (Fig.~\ref{fig:rank}), with lower numbers indicating that the evaluators more highly value the feature. For example, the average ranking of ``Students Performance Overview'' is 4.8, making it one of the most valued features. On the other hand, ``Role assignment suggestions within the group'' has an average ranking of 13.4, which is ranked lower by the evaluators.}

\begin{table*}[htbp]

\renewcommand{\arraystretch}{1.5}
\centering
\small
\begin{tabular}{l c c c c c c c c c c c c}
\hline
\textbf{Feature} & \textbf{T6} & \textbf{T5} & \textbf{E1} & \textbf{E2} & \textbf{TA1} & \textbf{TA2} & \textbf{T1} & \textbf{T2} & \textbf{T3} & \textbf{T4} & \textbf{Mean} & \textbf{SD} \\ \hline
F1 & 7 & 1 & 7 & 9 & 12 & 2 & 1 & 11 & 1 & 4 & 5.5 & 4.28 \\
F2 & 6 & 3 & 6 & 10 & 11 & 3 & 3 & 10 & 3 & 5 & 6 & 3.23 \\
F3 & 12 & 8 & 16 & 14 & 16 & 12 & 13 & 9 & 4 & 13 & 11.7 & 3.74 \\
F4 & 16 & 15 & 15 & 13 & 14 & 13 & 12 & 6 & 6 & 14 & 12.4 & 3.57 \\
F5 & 14 & 13 & 14 & 15 & 15 & 15 & 14 & 12 & 2 & 15 & 12.9 & 3.96 \\
F6 & 7 & 2 & 5 & 8 & 10 & 4 & 11 & 8 & 7 & 6 & 6.8 & 2.70 \\
F7 & 15 & 14 & 13 & 16 & 13 & 14 & 15 & 13 & 5 & 16 & 13.4 & 3.17 \\
F8 & 8 & 6 & 4 & 1 & 3 & 11 & 8 & 7 & 8 & 2 & 5.8 & 3.19 \\
F9 & 5 & 5 & 12 & 7 & 9 & 6 & 9 & 14 & 16 & 7 & 9 & 3.83 \\
F10 & 10 & 4 & 3 & 3 & 4 & 5 & 10 & 5 & 10 & 8 & 6.2 & 2.97 \\
F11 & 9 & 7 & 2 & 4 & 5 & 7 & 7 & 4 & 9 & 9 & 6.3 & 2.45 \\
F12 & 11 & 10 & 11 & 11 & 6 & 10 & 16 & 15 & 15 & 10 & 11.5 & 3.03 \\
F13 & 3 & 9 & 1 & 5 & 1 & 8 & 6 & 16 & 11 & 3 & 6.3 & 4.79 \\
F14 & 4 & 11 & 8 & 6 & 7 & 9 & 5 & 3 & 14 & 11 & 7.8 & 3.49 \\
F15 & 13 & 16 & 10 & 12 & 8 & 16 & 2 & 1 & 12 & 12 & 10.2 & 5.18 \\
F16 & 2 & 12 & 9 & 2 & 2 & 1 & 4 & 2 & 13 & 1 & 4.8 & 4.69 \\
\hline
\end{tabular}
\caption{Mean and SD Results of the Rankings for F1-F16 by Ten Participants.}
\label{table:rank}
\end{table*}

\begin{figure}[htbp]
    \centering
    \includegraphics[width=0.8\linewidth]{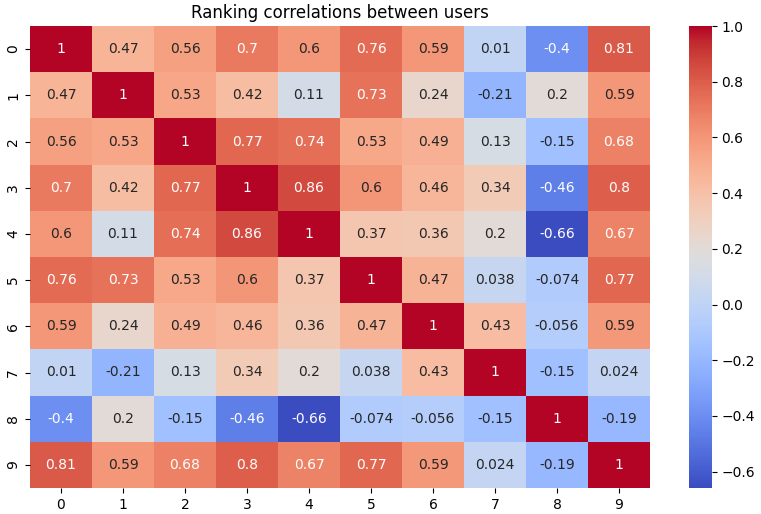}
    \caption{Spearman's Rank Correlation Results Between Participants.}
    \Description{Spearman's Rank Correlation Results Between Participants.}
    \label{fig:myplot}
\end{figure}

\RR{\textbf{Spearman Correlation Matrix:}
To assess the consistency between rankings, we used Spearman's Rank Correlation to calculate the correlation of rankings between each participant., The Spearman correlation matrix shows the consistency of the rankings between participants (Fig.~\ref{fig:myplot}). The correlation coefficient ranges from -1 to 1:
- A value closer to 1 indicates that the two evaluators' rankings are more consistent.
- A value closer to -1 suggests that the rankings are more inconsistent.
- A value near 0 implies no significant correlation between the two evaluators' rankings. For example, the correlation between participant 0 and participant 9 is 0.809, indicating
relatively consistent rankings. In contrast, the correlation between evaluator seven and evaluator 8 is - 0.150, suggesting their rankings differ considerably.}

\RR{\textbf{Friedman Test Results:}
We conducted a Friedman test to assess whether there are significant differences in rankings among different participants. The Friedman test compares the ranking data of multiple related samples to determine whether there are significant differences in rankings under different conditions. The test statistic is 60.796, and the p-value is 1.84e-07. This very small p-value (< 0.05) indicates that:

- There are significant differences in the rankings of different features. The evaluators' ratings are not random or similar but show clear preferences for specific features over others.

In conclusion, after comprehensively analyzing the results for each feature, our system focuses on implementing features \textbf{F1, F2, F6, F8-14, and F16}.
}
\newpage
\RR{
\section{PROMPTS}
Below are the prompts used to label data.
\subsection{Python Code Evaluation}
\textbf{Prompt:}

I would like you to play the role of a teacher who teaches a Python programming class, and you will be provided a question statement and a Python code, which is the student's answer to the question. 
Regarding the Python code, you need to accomplish two tasks. 
Here are the scoring criteria. Please mark each point according to the scoring criteria and explain the reason. Meanwhile, you should give your final score.
If the score of each aspect is not 5, please point out the demerits of the code. Also, note that you don't need to give the advised code. }

\begin{lstlisting}
criteria = {
    "Problem-solving Approach (5%)": {
        "Excellent (5)": "Shows an effective problem-solving approach, effectively addressing key challenges in the task",
        "Good (4)": "Shows a good problem-solving approach, with clear attempts to address challenges in the task",
        "Fair (3)": "Shows some effort in problem-solving, but lacks clarity or effectiveness in addressing challenges in the task",
        "Poor (2)": "Shows limited problem-solving efforts, with unclear or ineffective attempts to address challenges in the task",
        "Bad (1)": "Demonstrates no effective problem-solving approach, unable to address the task"
    },
    "Code Integrity (35%)": {
        "Excellent (5)": "The code is well-structured, organized, readable, and it effectively implements the desired functionality",
        "Good (4)": "The code structure and organization are sufficient, and it implements the function, though readability could be enhanced",
        "Fair (3)": "The code implements 80% of the function but lacks consistency or clarity",
        "Poor (2)": "The code implements 60% of the function but lacks consistency or clarity",
        "Bad (1)": "The code failed to implement 40% of the function and lacks consistency or clarity"
    },
    "Code Accuracy (35%)": {
        "Excellent (5)": "Code exhibits an excellent level of accuracy, producing correct results under various conditions and inputs",
        "Good (4)": "Code exhibits good accuracy, producing correct results under most conditions and inputs",
        "Fair (3)": "Code accuracy is acceptable, but there may be some occasional errors or inconsistencies in the results",
        "Poor (2)": "Code accuracy is poor, with frequent errors or inconsistencies in the results",
        "Bad (1)": "Code accuracy is severely lacking, with a high frequency of errors or inconsistencies in the results"
    },
    "Algorithm Innovation (25%)": {
        "Excellent (5)": "Shows innovative and creative approaches, showcasing originality and ingenuity in algorithm design and implementation",
        "Good (4)": "Shows basic innovation in algorithm design and implementation, but lacks originality or significant creativity",
        "Fair (3)": "Demonstrates limited innovation in algorithm design and implementation, with little originality or creativity demonstrated",
        "Poor (2)": "Shows little innovation in algorithm design and implementation, with no originality or creativity demonstrated",
        "Bad (1)": "Shows no innovation in algorithm design and implementation; only capable of imitating from simple examples"
    }
}
\end{lstlisting}

\RR{
\textbf{Task:} Task 1, you need to extract the key ideas of the code. Give me a paragraph outlining the thought process of this code. Task 2, Grade the code according to the scoring criteria, and don't round the final score.

\textbf{Input Example:}
}
\begin{lstlisting} 
Question: Existing list a=[49, 38 , 65 , 97 , 76 , 13 , 27 , 55 , 4]
Requirement: Try to write a program in Python that sorts the data elements in a from smallest to largest and prints out the new sorted list a.
Answer:
a=[49,38,65,97,76,13,27,55,4]
a.sort()
print(a)       
\end{lstlisting} 
\RR{
\textbf{Output Example:}}
\begin{lstlisting}
{
    "Key ideas": "The provided code aims to sort a given list of integers in ascending order. The process follows these key steps:\n\t1. Define a list named `a` containing a series of integers.\n\t2. Call the `sort()` method on the list `a`, which sorts the list in place in ascending order.\n\t3. Print the sorted list to display the numbers from smallest to largest.\nThis approach efficiently utilizes Python's built-in sorting capabilities and achieves the task as specified in the question.",
    "Score": "4.55 / 5",
    "Details": [
        {
            "Problem-solving Approach (5%)": {
                "Score": "Excellent (5)",
                "Explanation": "The code presents a clear and effective approach to sorting the list using Python's built-in functionality, which is suitable for the problem at hand."
            },
            "Demerits": null
        },
        {
            "Code Integrity (35%)": {
                "Score": "Excellent (5)",
                "Explanation": "The code is well-structured, organized, and highly readable. It makes effective use of the `sort()` method while maintaining clarity. There are no unnecessary complexities in the implementation."
            },
            "Demerits": null
        },
        {
            "Code Accuracy (35%)": {
                "Score": "Excellent (5)",
                "Explanation": "The code accurately sorts the list `a` and produces the correct output without any errors. The use of the `sort()` method guarantees that the order will be correct."
            },
            "Demerits": null
        },
        {
            "Algorithm Innovation (25%)": {
                "Score": "Fair (3)",
                "Explanation": "While the implementation is effective and clear, it does not demonstrate any innovative or creative algorithmic approach since it relies on the built-in `sort()` method without any modifications or enhancements."
            },
            "Demerits": "There is a lack of originality in the sorting technique, as it could have showcased a custom sorting algorithm, which would demonstrate deeper understanding and application of algorithms."
        }
    ]
}
\end{lstlisting} 
\RR{\subsection{Collaborative Behavioral Analysis}
\textbf{Prompt:}}
\RR{
You are a teacher teaching programming class and will be provided some conversation files belonging to one group but in different questions period, including question number, timestamp for each speaker and corresponding conversation recorded in the context of a collaborative programming course. It will be your job to find students' communication behaviors for all sentences they said and record them in a JSON format file, besides, show the prediction percentage of your response after each behavior by using the scaffold.
Notice again, you should show all the sentences in the JSON file. If you can't classify some of them into category, indicate them into one category with the most similar meaning and point out the percentage. 
Specifically, each question is encircled by a ``{ },'' and you need to show details information in it.

\textbf{Output Format:}}
\begin{lstlisting}
{
    "Question": "Question Number",
    "Conversations": [
        {
            "Speaker": "",
            "Timestamp": "",
            "Content": "",
            "Behavior Category": "",
            "Prediction Percentage": "",
            "Explanation": ""
        },
        {
            "Speaker": "",
            "Timestamp": "",
            "Content": "",
            "Behavior Category": "",
            "Prediction Percentage": "",
            "Explanation": ""
        }
    ]
}
\end{lstlisting}

\RR{
\textbf{Input Example:}}
\begin{lstlisting}
{
    "role": "user",
    "content": """
        Question1
        1.00 2.90 0303 This one is better done
        2.90 13.90 0302 This one compares This first question is so good that it doesn't need doesn't need to be discussed too much, what else could be wrong with it? There's really no problem with it.
        13.90 14.90 0303 Don't don't laugh eh?
        21.00 24.30 0303 Good, the first question is not too difficult, successfully completed
        34.30 40.40 0302 Whatever, just mess around with the name, just send this when the time comes.
        40.40 43.40 0302 Good success, perfect, success.
         """
}
\end{lstlisting}

\RR{
\textbf{Output Example:}}
\begin{lstlisting}
{
    "Question": "Question1",
    "Conversations": [
        {
            "Speaker": "0303",
            "Timestamp": "1.00-2.90",
            "Content": "This is better done",
            "Behavior Category": "Project understanding",
            "Prediction Percentage": "100%",
            "Explanation": "Member expresses that the task is easy, indicating understanding of the project."
        },
        {
            "Speaker": "0302",
            "Timestamp": "2.90-13.90",
            "Content": "This comparison of this first question is so good that it doesn't need to be without too much discussion, what could possibly be wrong with it? There really is no problem with it.",
            "Behavior Category": "Project understanding",
            "Prediction Percentage": "90%",
            "Explanation": "Member emphasizes that the first task is simple and questions if there are any concerns, showing clarity in project understanding."
        },
        {
            "Speaker": "0303",
            "Timestamp": "13.90-14.90",
            "Content": "Don't laugh.",
            "Behavior Category": "Unrelated chat",
            "Prediction Percentage": "60%",
            "Explanation": "This is unrelated humor, not pertinent to the task, indicating a digression from the activity."
        },
        {
            "Speaker": "0303",
            "Timestamp": "21.00-24.30",
            "Content": "Okay, the first question was not too difficult and went well",
            "Behavior Category": "Acknowledgement",
            "Prediction Percentage": "100%",
            "Explanation": "The member acknowledges successful completion of the first task."
        },
        {
            "Speaker": "0302",
            "Timestamp": "34.30-40.40",
            "Content": "Whatever, just make up a name, and when the time comes, just send this.",
            "Behavior Category": "Unrelated chat",
            "Prediction Percentage": "50%",
            "Explanation": "Casual discussion about naming, which strays from focused collaboration."
        },
        {
            "Speaker": "0302",
            "Timestamp": "40.40-43.40",
            "Content": "So successful, perfect, successful.",
            "Behavior Category": "Acknowledgement",
            "Prediction Percentage": "80%",
            "Explanation": "Member acknowledges success with positive feedback."
        }
    ]
}
\end{lstlisting}
\RR{
\subsection{Student Role Analysis}
\textbf{Prompt:}}

\RR{
You are a teacher teaching programming class and will be provided some conversation files belonging to one group but in different questions period, including question number, timestamp for each speaker and corresponding conversation recorded in the context of a collaborative programming course. Your jobYour job will be to find students' planning solutions behaviors related to the question for all sentences they said. 
Here is some example: ``This question could go like this...'', ``Combine A and B'', ``Notice the function.'', etc. Specifically, you should indicate each speaker's sentences of planning solutions corresponding timestamp. Besides, there are some misunderstanding sentences that are not planning solutions: ``I'm just messing around with names.'', ``That's good. That's good.'' etc. You need to find valuable comments that contribute or drive the problem-solving process. Notice that you only need to indicate the sentences about planning solutions or providing insights.
Let me explain ``Navigator'', ``Driver'' and ``Monitor''. Each group only have three members, ``Navigator'' is the speaker who's sentence is about planning solutions, ``Driver'' is the member who responsible for coding, this role is non-changeable. I will tell you who are the ``Driver'' in each file's first line. Besides, ``Monitor'' is the role who is neither 
``Navigator'' nor ``Driver''. If the role of ``Driver'' is also planning solutions, then the role of ``Driver'' should be changed to ``Navigator'', and ``Driver'' is None for this sentence. Significantly, if a sentence is not about planning solutions, you also need to list/indicate it. The Navigator should be None, but fill out the speakers of ``Monitors'' and ``Drivers''.

\textbf{Output Format:}}
\begin{lstlisting}
{
    "Question": "QuestionX",
    "Conversations": [
        {
            "Timestamp": "XXX",
            "Content": "XXX",
            "Navigator": "XXX",
            "Other_Roles": [
                {
                    "Monitors": ["XXX"],
                    "Driver": ["XXX"]
                }
            ]
        },
        {
            "Timestamp": "XXX",
            "Content": "XXX",
            "Navigator": "XXX",
            "Other_Roles": [
                {
                    "Monitors": ["XXX"],
                    "Driver": ["XXX"]
                }
            ]
        }
    ]
}
\end{lstlisting}

\RR{
\textbf{Input Example:}}
\begin{lstlisting}
"""
    Question2 Driver: 0302
    46.70 49.40 0303 Question two, question two, continue with question two.
    58.10 60.90 0302 What is the title of the second question?
    60.90 62.20 0303, and the question is.
    75.20 77.90 0301, also in descending order.
    78.90 80.80 0302 That's okay; that should be the template for the first question.
    80.80 81.80 0301 AB merge and then
    81.80 87.30 0303 Right, right, the second question would also work, yes
    91.30 98.50 0302 Right, this one should be noted that this ijk he this exclamation point should be an introduction to the.
"""
\end{lstlisting}

\RR{
\textbf{Output Example:}}
\begin{lstlisting}
{
    "Question": "Question2",
    "Conversations": [
        {
            "Timestamp": "46.70-49.40",
            "Content": "Question two. Question two. Go on to question two.",
            "Navigator": "0303",
            "Other_Roles": [
                {
                    "Monitors": ["0301"],
                    "Drivers": ["0302"]
                }
            ]
        },
        {
            "Timestamp": "58.10-60.90",
            "Content": "What is the title of the second question?",
            "Navigator": "0302",
            "Other_Roles": [
                {
                    "Monitors": ["0301", "0303"],
                    "Drivers": ["None"]
                }
            ]
        },
        {
            "Timestamp": "60.90-62.20",
            "Content": "The question is.",
            "Navigator": "0303",
            "Other_Roles": [
                {
                    "Monitors": ["0301"],
                    "Drivers": ["0302"]
                }
            ]
        },
        {
            "Timestamp": "75.20-77.90",
            "Content": "It's also sorted from childhood to adulthood.",
            "Navigator": "0301",
            "Other_Roles": [
                {
                    "Monitors": ["0303"],
                    "Drivers": ["0302"]
                }
            ]
        },
        {
            "Timestamp": "78.90-80.80",
            "Content": "That's okay. That would be the template for the first question.",
            "Navigator": "0302",
            "Other_Roles": [
                {
                    "Monitors": ["0301", "0303"],
                    "Drivers": ["None"]
                }
            ]
        },
        {
            "Timestamp": "80.80-81.80",
            "Content": "AB is merged and then",
            "Navigator": "0301",
            "Other_Roles": [
                {
                    "Monitors": ["0303"],
                    "Drivers": ["0302"]
                }
            ]
        },
        {
            "Timestamp": "81.80-87.30",
            "Content": "Yeah, yeah, the second question would work, too. Yeah.",
            "Navigator": "0303",
            "Other_Roles": [
                {
                    "Monitors": ["0301"],
                    "Drivers": ["0302"]
                }
            ]
        },
        {
            "Timestamp": "91.30-98.50",
            "Content": "Yes, this one should be noted that this ijk he this exclamation point is supposed to be an introduction.",
            "Navigator": "0302",
            "Other_Roles": [
                {
                    "Monitors": ["0301", "0303"],
                    "Drivers": ["None"]
                }
            ]
        }
    ]
}
\end{lstlisting}

\RR{
\subsection{Teacher Scaffold Analysis}

\textbf{Prompt:}

You are a teacher teaching programming class and you will be provided a conversation file including timestamp for each speaker and corresponding content recorded in the context of a collaborative programming course. It will be your job to find instructors' assitance category based on the following scaffold.
Analyze the different levels of scaffolding used by instructors during group learning based on the following categories:
Low-control cognitive scaffolding (CS-L): The instructor raises open-ended questions that elicit group thinking without providing new information. This method encourages critical thinking but leaves the group to figure out the details.      
Medium-control cognitive scaffolding (CS-M): The instructor provides hints or clues to help groups solve cognitive problems. This method supports problem-solving but maintains some cognitive challenge.       
High-control cognitive scaffolding (CS-H): The instructor directly provides answers or demonstrates tasks (such as programming) using tools like computers. This method offers direct guidance but may limit students' independent problem-solving.      
Metacognitive scaffolding (MS): The instructor monitors and regulates the group's learning goals and collaborative processes, helping to manage group dynamics and learning strategies.

\textbf{Input Example:}}
\begin{lstlisting}
"""
    131.0 133.0 0000 What's the problem? Oh, I see. 
    134.0 135.0 0000 3 Hmm, yes, exactly. 
    139.0 142.0 0000 Hmm, I, I, where is that program? 
    143.0 196.0 0000 Then here, you need to indent first, right? Wherever you don't finish, you must add a colon, okay? Hmm, like in the if statement, you're correct about that. I, I, it should be I != K, right? Hmm, and and means "and", K should not be equal to J, right? And another one, I should also not be equal to K, right? Isn't it true that I, J, and K should all be different from each other? So if this condition is met, what happens next? Then you calculate something, calculate an S. The S would be equal to I * 100 + J * 10 + K, right? Is that correct? But don't forget to print it out, so print what? print this S, and that's it.
"""
\end{lstlisting}
\RR{
\textbf{Output Example:}}

\begin{lstlisting}
{
    "Speaker": "0000",
    "Timestamp": "131.0-133.0",
    "Content": "What is the question? Oh, oh, yes.",
    "Behavior Category": "Metacognitive scaffolding",
    "Prediction Percentage": "80%",
    "Explanation": "Requests clarification about the third question.",
},
{
    "Speaker": "0000",
    "Timestamp": "134.0-135.0",
    "Content": "Three digits, yes, yes.",
    "Behavior Category": "Metacognitive scaffolding",
    "Prediction Percentage": "90%",
    "Explanation": "Acknowledges the provided information.",
},
{
    "Speaker": "0000",
    "Timestamp": "139.0-142.0",
    "Content": "Hmm, where is your program? Oh.",
    "Behavior Category": "Metacognitive scaffolding",
    "Prediction Percentage": "90%",
    "Explanation": "Teacher responds by asking about the student's program location.",
},
{
    "Speaker": "0000",
    "Timestamp": "143.0-196.0",
    "Content": "Then you start with if here, make sure to indent, and always use a colon after each if. Yes, if i is not equal to k, right? And also k is not equal to j. I also can't be equal to k, right? So, if i, j, and k are not equal to each other, then what? You calculate something, calculate an s, which equals i times 100, plus j times 10, plus k. Right? But you have to remember to print it out. Print s, and that's it.",
    "Behavior Category": "High-control cognitive scaffolding",
    "Prediction Percentage": "100%",
    "Explanation": "Teacher provides detailed explanation and coding instructions.",
}
         
\end{lstlisting}

\end{document}